\documentclass[aps,prx,twocolumn,superscriptaddress,longbibliography]{revtex4-2}
\usepackage{amsmath,amssymb}
\usepackage[pdftex]{hyperref,graphicx}
\hypersetup{colorlinks = true, urlcolor = blue, linkcolor = blue, citecolor = blue}
\usepackage{physics}
\usepackage{xcolor}
\usepackage{bm}
\usepackage[normalem]{ulem}

\newcommand{\para}[1]{}

\begin{document}

\author{Haining Pan}
\affiliation{Department of Physics and Astronomy, Center for Materials Theory, Rutgers University, Piscataway, NJ 08854 USA}
\author{Nakul Aggarwal}
\affiliation{Department of Physics and Astronomy, Center for Materials Theory, Rutgers University, Piscataway, NJ 08854 USA}
\author{J. H. Pixley}
\affiliation{Department of Physics and Astronomy, Center for Materials Theory, Rutgers University, Piscataway, NJ 08854 USA}
\affiliation{Center for Computational Quantum Physics, Flatiron Institute, New York, New York 10010, USA}

\begin{abstract}

\end{abstract}
\title{
Pruning-induced phases in fully-connected neural networks: the eumentia, the dementia, and the amentia 
}

\begin{abstract}
    Modern neural networks are heavily overparameterized, and pruning, which removes redundant neurons or connections, has emerged as a key approach to compressing them without sacrificing performance. However, while practical pruning methods are well developed, whether pruning induces sharp phase transitions in the neural networks and, if so, to what universality class they belong, remain open questions. To address this, we study fully-connected neural networks trained on MNIST, independently varying the dropout (i.e., removing neurons) rate at both the training and evaluation stages to map the phase diagram. We identify three distinct phases: eumentia (the network learns), dementia (the network has forgotten), and amentia (the network cannot learn), sharply distinguished by the power-law scaling of the cross-entropy loss with the training dataset size. {In the eumentia phase, the algebraic decay of the loss, as documented in the machine learning literature as neural scaling laws, is from the perspective of statistical mechanics the hallmark of quasi-long-range order.} We demonstrate that the transition between the eumentia and dementia phases is accompanied by scale invariance, with a diverging length scale that exhibits hallmarks of a Berezinskii-Kosterlitz-Thouless-like transition; the phase structure is robust across different network widths and depths. Our results establish that dropout-induced pruning provides a concrete setting in which neural network behavior can be understood through the lens of statistical mechanics.
\end{abstract}

\maketitle

% \tableofcontents
\section{Introduction}
\para{History: NN are overparameterized, many examples, like LLM andrey gromov "The Unreasonable Ineffectiveness of the Deeper Layers", and other papers -- this is a very mature topic so it should be easy to find many examples.}
Modern neural networks are heavily overparameterized, containing far more trainable parameters than needed to fit the training data.
For example, large language models, a class of neural networks with billions of parameters~\cite{brown2020language, touvron2023llama, chowdhery2023palm}, have a significant fraction of deeper layers that can be removed with negligible performance degradation~\cite{gromov2024unreasonable}. A similar layer-wise and parameter-wise redundancy has been identified in convolutional networks~\cite{han2015learning,li2017pruning}, transformers~\cite{men2025shortgpt}, and other architectures~\cite{hoefler2021sparsity}. Understanding the origin and role of this overparameterization remains a central question in modern deep learning, and delivering smaller models is also practically desirable, as reduced model size translates directly to lower memory, faster inference, and more economical deployment.

\begin{figure*}[htbp]
    \centering
    \includegraphics[width=7in]{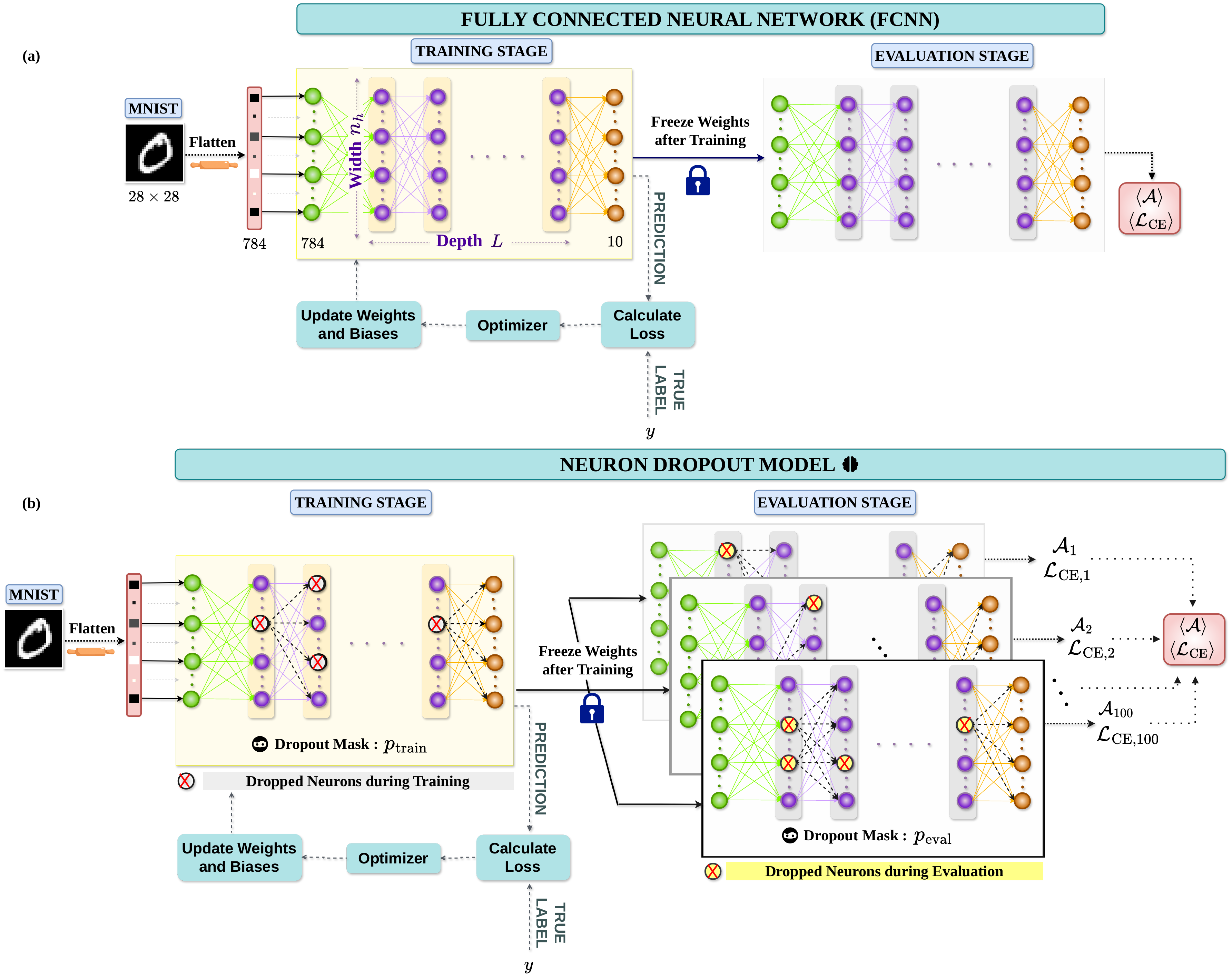}
    \caption{
        (a) Schematic of the fully-connected neural network. A $28\times 28$ gray scale image from MNIST dataset is first flattened into a $784$-dimensional input vector, and then passed through a rectangular architecture of depth $L$ hidden layers and width $n_h$, where each hidden layer contains $n_h$ neurons. During training, the weights and biases are optimized on the training set by minimizing the cross-entropy loss using the Adam optimizer. 
        During evaluation, these optimized parameters are frozen, and the network is evaluated to compute performance metrics including the classification accuracy $\langle\mathcal{A}\rangle$ and cross-entropy loss $\langle\mathcal{L}_{\text{CE}}\rangle$ on test dataset. 
        The average values are obtained over $20$ independent training runs with different initializations and mini-batch samplings. (See Sec.~\ref{sec:model} for details.)
        (b) Schematic of the neuron-dropout model. The network architecture is identical to that in panel (a), except neurons in the hidden layers are randomly masked by setting their post-activation outputs to zero. The neurons are dropped independently with dropout rates $p_{\mathrm{train}}$ and $p_{\mathrm{eval}}$ during training and evaluation stages respectively. For each fixed $p_{\text{eval}}$, we sample $100$ independent dropout masks and estimate the mean accuracy.
    }
    \label{fig:schematic}
\end{figure*}

\para{Pruning as a way to reduce overparameterization, and relevant works including different ways of pruning chronologically: OBD/OBS (classical, weight-level) $\rightarrow$  magnitude (modern, weight-level) $\rightarrow$  structured (neuron/filter/layer-level) $\rightarrow$  stochastic/random (dropout, dropconnect).}
One natural approach to reducing overparameterization is \textit{pruning}, which removes components from a trained network to produce a smaller, sparser model.
The simplest pruning strategy removes individual weights with the smallest absolute values. To overcome the shortcomings of this naive magnitude-based approach, Optimal Brain Damage~\cite{lecun1989optimal} and Optimal Brain Surgeon~\cite{hassibi1993optimal} instead use the second-order sensitivity (saliency) of the loss function to identify which weights to remove. In the modern deep learning era, Han et al.~\cite{han2015learning} showed that magnitude pruning, when combined with iterative retraining, achieves effective compression of deep networks. 
However, these weight-level methods produce irregular sparse matrices that are difficult to accelerate on standard hardware. Structured pruning addresses this by permanently removing entire rows or columns of the weight matrices---equivalently, removing all incoming and outgoing connections of selected neurons, or even entire layers---yielding a smaller but dense network that achieves direct speedup without specialized sparse computation~\cite{wen2016learning, li2017pruning}.
All of the above methods are deterministic, selecting which components to remove based on scores derived from the trained weights. Separately, dropout~\cite{srivastava2014dropout} (see Fig.~\ref{fig:schematic}(b)) and its generalization dropconnect~\cite{wan2013regularization} were originally introduced as stochastic regularization techniques that temporarily deactivate neurons or connections at random during training to prevent co-adaptation. These methods can also be interpreted as forms of random pruning: dropout at the neuron level and dropconnect at the connection level.
% \jp{JP: Reference Fig 1 as needed throughout above paragraph}

\para{Fundamentally, a conjecture called the lottery ticket hypothesis is that there exists a subnetwork that can be trained effectively, and the pruning process is to find such a subnetwork. which justifies why we need large NN , why it is beneficial in the training stage, but in the evaluation stage, i.e., when we deliver the model, we can reduce size.}
A fundamental question underlying all pruning methods is why overparameterized networks can be compressed so aggressively. The lottery ticket hypothesis (LTH)~\cite{frankle2018lottery} offers a compelling conjecture: a large, randomly initialized network contains sparse subnetworks, ``winning tickets'' that, when trained in isolation, can match the performance of the full network. In this view, overparameterization is not wasteful but rather increases the probability of containing such a trainable subnetwork~\cite{malach2020proving}, while also smoothing the loss landscape so that gradient descent reliably converges to good solutions~\cite{du2019gradient, allen-zhu2019convergence}. Once training is complete and the winning ticket has been identified through pruning, the redundant parameters can be discarded, yielding a compact model for deployment.

\para{The lottery ticket remain at the level of conjectures and empirical optimizationremainsnomenological in the physics sense---without a theory. Cite lots of sparsity paper. We try to provide a physicist perspective: are there phase transitions? what universality class? can we write down an effective model? To study phase transitions, we need a continuously tunable control parameter, and the dropout rate gives us exactly that.}
Despite being simple and compelling, the LTH remains a hypothesis, and a more fundamental understanding of pruning remains an open question. While many papers have focused on developing and benchmarking pruning algorithms from a practical standpoint~\cite{zhu2018prune, liu2018rethinking, zhang2024how, evci2020rigging,hoefler2021sparsity}, comparatively little attention has been paid to the fundamental ``physics'' of pruning. 
The LTH implies that a mildly pruned network should hardly affect the performance, while, intuitively, violent pruning leads to the entire breakdown of the network's function, {and the more extreme limit of a disconnected network simply cannot work at all. 
This motivates us to bring a physicist's perspective to the problem, we can view the regime when the network can be trained as a stable ``phase'' that is robust to small amounts of pruning, whereas the limit of large pruning indicates a distinct phase that can no longer remember what it has learned. A fundamental question then becomes,
as the pruning rate increases, does the breakdown occur as a smooth crossover or through a sharp phase transition between these two regimes that exhibits scale invariance? If the latter, what universality class does it belong to?} Can the critical behavior be captured by an effective statistical mechanics model? The dropout rate provides a natural starting point for such a study, as it serves as a continuously tunable control parameter.

\para{So that motivates us to look at two different types of dropout. Especially the novelty is that evaluation dropout is rare in delivery. We also look at ``random pruning'' compared to ML community which is often magnitude pruning. Especially we use neuron dropout (not drop connect) to avoid a trivial percolation transition by design.}
In standard practice, dropout is applied only during training and turned off at evaluation, i.e., when the trained network is applied to unseen data~\cite{srivastava2014dropout}. From a physics standpoint, however, varying the dropout rate at evaluation is equally natural because it probes how robust the trained network is to the removal of neurons. While evaluation dropout has been explored for uncertainty estimation~\cite{gal2016dropout}, independently varying it as a control parameter separate from the training dropout rate has not been systematically studied. By independently varying the dropout rate at both stages, the training dropout rate and the evaluation dropout rate, we map the phase diagram in the two-dimensional plane.
Since dropout removes neurons irrespective of their learned values, the dropout rate serves as a clean external tuning knob independent of the network's state, in contrast to magnitude-based pruning, where the pruning pattern is coupled to the learned weights. Specifically, we employ neuron dropout rather than dropconnect~\cite{wan2013regularization} to avoid the trivial percolation limit that arises from random removal of individual connections to disconnect the output from the input.

\para{For simplicity, we start with the most basic architecture, the fully-connected neural network (FCNN). We vary the training dropout rate and evaluation dropout rate, and find three phases---eumentia, dementia, and amentia---characterized by the trainability and generalization of the network. Define them and make the brain analogy.}
To investigate these questions, we start with the simplest architecture, the FCNN, trained on the MNIST handwritten digit dataset. The training dropout rate controls whether the network can learn from the data, while the evaluation dropout rate probes whether the trained network can still generalize when neurons are removed. We find that their independent variation gives rise to three distinct phases, which are sharply distinguished by how the test cross-entropy loss scales when we vary the amount of training data. 
Since dropout effectively removes neurons from the network, a natural analogy to clinical neuroscience suggests itself: we call these the \emph{eumentia}, \emph{dementia}, and \emph{amentia} phases. In the eumentia phase, the network learns and generalizes effectively, where the loss decreases with more data, analogous to a healthy brain. In the dementia phase, the network trains successfully but loses its ability to generalize when neurons are removed at evaluation: the loss actually increases with more data, reminiscent of cognitive decline. In the amentia phase, excessive training dropout prevents the network from learning altogether; the loss is insensitive to the amount of data, analogous to a brain that never developed the capacity to learn at birth. {Across all three phases, the cross-entropy loss scales as a power law with the training dataset size. In the eumentia phase, the loss decays with more data, consistent with the neural scaling laws documented in the machine learning literature~\cite{seung1992statistical, hestness2017deep, spigler2020asymptotic, kaplan2020scaling, rosenfeld2020constructive, bahri2020statistical,bordelon2020spectrum}. From a physics standpoint, such algebraic decay is the hallmark of QLRO.} We further find that the transition between the eumentia and dementia phases suggests a Berezinskii-Kosterlitz-Thouless (BKT) like transition, supported by finite-size scaling analysis.

\para{In the following section}
The rest of the paper is organized as follows. In Sec.~\ref{sec:model}, we define the FCNN architecture, the training and evaluation protocol, and the implementation of the two dropout rates. In Sec.~\ref{sec:results}, we present the phase diagram, characterize the three phases through the scaling of the test cross-entropy loss with the training dataset size, analyze the BKT-like transition between the eumentia and dementia phases using finite-size scaling, examine the classification accuracy as a complementary diagnostic, and demonstrate that the phase structure is robust across different network widths and depths. We conclude with a summary and outlook in Sec.~\ref{eq:conclusion}. Appendices~\ref{app:bkt} and \ref{app:powerlaw} detail the finite-size scaling analyses using the BKT-like and conventional algebraic fits respectively. Appendix~\ref{app:MNIST} describes the MNIST dataset, and Appendix~\ref{app:dropout} provides technical details of the neuron dropout implementation.
% \para{We focus on T goes to inf limit, intuitively speaking, by using Adam to study the steady state phase diagram.}

\section{Model and Methods}\label{sec:model}
We now turn to describing the neural network and pruning models that we investigate. In the following, we define and discuss the cross entropy and the accuracy of the neural network. 

\subsection{fully-connected neural networks}\label{sec:fcnn}
\para{We consider FCNN, define the structure and layer, specify the numbers, also we need to mention, we study FCNN with depth up to 5 since the deeper network is harder to train and cite the ResNet paper.}
We consider an FCNN with $L$ hidden layers, each with the same width $n_h$ as shown in Fig.~\ref{fig:schematic}(a).
We choose the depth of networks (number of hidden layers) $L$ from 2 to 5 to avoid the vanishing gradient issue in training the deep FCNN without using skip connections (i.e., dropconnect)~\cite{bengio1994learning, hochreiter1998vanishing, glorot2010understanding, he2016deep}, and the width $n_h$ from 50 to 200. We find that the detailed dimension of the FCNN does not qualitatively change the phase diagram.
% \jp{JP: Refer to Figure 1 as needed in this section.}

\para{Input and output dimension. Mention that we use MNIST, use one sentence to briefly explain what MNIST is, and then refer to Appendix~\ref{app:MNIST}}
We use the MNIST dataset~\cite{lecun1998gradientbased} as the training dataset (see Appendix~\ref{app:MNIST}), which is a benchmark of 70000 handwritten digit images from 0 to 9. 
The input layer $\bm{x}$ is a $784$-dimensional vector, flattened from the original $28\times 28$ pixel grayscale image, and the output layer has $10$ units corresponding to the $10$ digit classes, as shown in the leftmost and rightmost layers of the FCNN in Fig.~\ref{fig:schematic}.

\para{Activation functions explain.}
For hidden layer $l = 1, \dots, L$, the activations are
\begin{equation}
h^{(l)} = \phi\left(W^{(l)} h^{(l-1)} + b^{(l)}\right),
\label{eq:fcnn_layer}
\end{equation}
where $h^{(0)} \equiv \bm{x}$, $W^{(l)} \in \mathbb{R}^{n_h \times n_h}$ and $b^{(l)} \in \mathbb{R}^{n_h}$ are trainable weights and biases (except for $W^{(1)} \in \mathbb{R}^{n_h \times 784}$), and $\phi(z) = \max(0, z)$ is the rectified linear unit (ReLU) activation function. The output layer has no activation and produces unnormalized logits $h^{(L+1)} = W^{(L+1)} h^{(L)} + b^{(L+1)}$, where $W^{(L+1)} \in \mathbb{R}^{10 \times n_h}$.

\para{Explain training/evaluation stage, and data split; Explain Accuracy and cross entropy, say that these are the two main metric we are looking at in this paper.}
We use the standard train/test split, with a fixed test set of $N_{\text{test}}=10000$. 
From the training set, we draw a subset of $N=\left\{ 10000, 20000, 30000, 40000, 50000, 60000 \right\}$ samples to study the effect of dataset size, and reserve $10\%$ for validation used in early stopping.
We follow standard training practice: the Adam optimizer~\cite{kingma2015adam} with a learning rate of $10^{-3}$, no weight decay, and a plateau-based learning rate scheduler that halves the learning rate when the validation loss stagnates for $10$ epochs. The early stopping is based on the validation loss with patience $20$, i.e., training is terminated if the validation loss does not improve for $20$ consecutive epochs. {Viewing this process from a physical system perspective, training epoch plays the role of time ($t$), and we can thus interpret the results that we present in the following as corresponding to the $t\to\infty$ limit. It will be interesting to study this ``dynamics'' in future work.}

In the training stage, the training objective is the cross-entropy loss $\mathcal{L}_{\text{CE}}^T=-\frac{1}{B}\sum_{b=1}^{B} \log p_{y^{(b)}}^{(b)}$ over a mini-batch of $B=64$ samples, where $y^{(b)}$ is the true class label of sample $b$ and the predicted class probabilities are obtained via softmax $p_i^{(b)} = e^{h_i^{(L+1)}} / \sum_{j=1}^{10} e^{h_j^{(L+1)}}$.

In the evaluation stage, we characterize network performance using two metrics. The cross-entropy loss evaluated on the full test set,
\begin{equation}
\mathcal{L}_{\text{CE}} = -\frac{1}{N_{\text{test}}}\sum_{m=1}^{N_{\text{test}}} \log p_{y^{(m)}}^{(m)},
\end{equation}
and the classification accuracy,
\begin{equation}\label{eq:A}
\mathcal{A} = \frac{1}{N_{\text{test}}}\sum_{m=1}^{N_{\text{test}}} \mathbf{1}\left(\hat{y}^{(m)} = y^{(m)}\right),
\end{equation}
where $\hat{y}^{(m)} = \arg\max_i h_i^{(L+1)}$ is the predicted label and $\mathbf{1}(\cdot)$ is the indicator function.
{Because the indicator function discards prediction confidence, for example, a sample correct at $99\%$ and one barely correct at $51\%$ contribute equally to $\mathcal{A}$ despite vastly different cross-entropy losses, the cross-entropy loss is a finer probe of the network's state and serves as our primary diagnostic throughout this work.}
Finally, averaging over the stochasticity from weight initialization, mini-batch sampling, and dropout masking in both the training and evaluation stages, we study the averaged cross-entropy loss $\expval{\mathcal{L}_{\text{CE}}}$ and accuracy $\expval{\mathcal{A}}$, where $\expval{\cdot}$ denotes the average over $20$ independent training runs and $100$ evaluation forward passes per trained model.

%\jp{JP: Is it meaningful to discuss (i) the connection between CE and free energy? (ii) The issue with accuracy using the indicator function?}\
% \HPcom{Both are good points. I added the second point as they are intuitive to address. TODO: for the first point, i will do some survey first and try to make it rigorous. }
%\HPcom{Update on (i): The CE loss decomposes as $\mathcal{L}_{\text{CE}} = -\log p_y = -h_y + \log \sum_j e^{h_j}$. The softmax is a Boltzmann distribution at $\beta=1$ with logits as negative energies, so the second term is the log-partition function $\log Z = -F$. This gives $\mathcal{L}_{\text{CE}} = E_y - F$, i.e., the energy of the correct class minus the free energy. So CE is not the free energy itself but the surprisal of the correct class, which mixes the energy and free energy. Without a rigorous theory connecting our phase transitions to a thermodynamic potential, I think elaborating on this analogy in the paper risks overstating a structural observation. I'd prefer to leave it out. (ii) is actually added.}

% explain the randomness in the training and testing phase in both train and evaluation stage
% We draw $N$ samples from the MNIST training set of $60000$ images and evaluate on the held-out test set of $10000$ images. 

\subsection{Training and evaluation dropout rate}\label{sec:dropout}
\para{Dropout is a way to prevent overfitting and co-adaptation}
In order to drive the phase transitions in the FCNN, we apply neuron dropout~\cite{srivastava2014dropout} to the hidden layers, which is a standard regularization technique to suppress co-adaptation among neurons and reduce overfitting. 
Neuron dropout means that a random subset of neurons in each hidden layer is temporarily deactivated during training by setting its output $h_i^{(l)} = 0$, effectively zeroing out all outgoing connections from that neuron to layer $l+1$ as shown in Fig.~\ref{fig:schematic}(b).
% ~\HPcom{Nakul, i think we need to deliver this message in the schematic.}
{A fresh dropout mask is independently sampled on every forward pass during both training and evaluation: no mask persists across iterations or from the training stage to the evaluation stage. The training and evaluation dropout are therefore two entirely independent stochastic processes applied to the same learned network parameters.}
% \jp{JP: Reference Fig 1 as needed}
Conventionally, neuron dropout is applied only during training and turned off at evaluation~\cite{srivastava2014dropout}. 
Applying dropout at the evaluation stage has been used to estimate model uncertainty~\cite{gal2016dropout}, 
in contrast, in the following work we use this form of drop out to study the stability and instability of the trained network.
%explored in the context of uncertainty estimation via Monte Carlo (MC) dropout~\cite{gal2016dropout}; however, uncertainty estimate is not our main point in this paper. 
%
Instead, we apply dropout at both training and evaluation (Fig.~\ref{fig:schematic}(b)), and independently vary the two rates to study phase transitions in the $(p_{\text{train}}, p_{\text{eval}})$ plane induced by dropout. For more details on the dropout implementation and rescaling of activations, see App. \ref{app:dropout}.

% \para{Explain the two dropout rate and show the mask math}
% For each hidden layer ($l = 1, \dots, L$), a binary mask is independently sampled for every neuron:
% \begin{equation}
% m_i \sim \text{Bernoulli}(1 - p),
% \end{equation}
% where $m_i = 1$ retains neuron $i$ and $m_i = 0$ deactivates it. During training, we apply dropout with rate $p = p_{\text{train}}$ and optimize the model parameters. After training, the learned parameters are fixed. We then apply dropout during evaluation with rate $p = p_{\text{eval}}$, and study network performance across the $(p_{\text{train}}, p_{\text{eval}})$ plane.

% \para{Rescale part}
% We use inverted dropout, where surviving activations are rescaled to preserve their expected magnitude:
% \begin{equation}
% \tilde{h}_i = \frac{m_i}{1 - p} h_i,
% \end{equation}
% so that $\mathbb{E}[\tilde{h}_i] = h_i$, ensuring that the scale of activations remains unchanged between training and evaluation.
\section{Results}\label{sec:results}
In the following sections, we begin with a summary of the phase diagram, and then present the supporting numerical evidence.

\subsection{Phase diagram}

\begin{figure*}[htbp]
    \centering
    \includegraphics[width=7in]{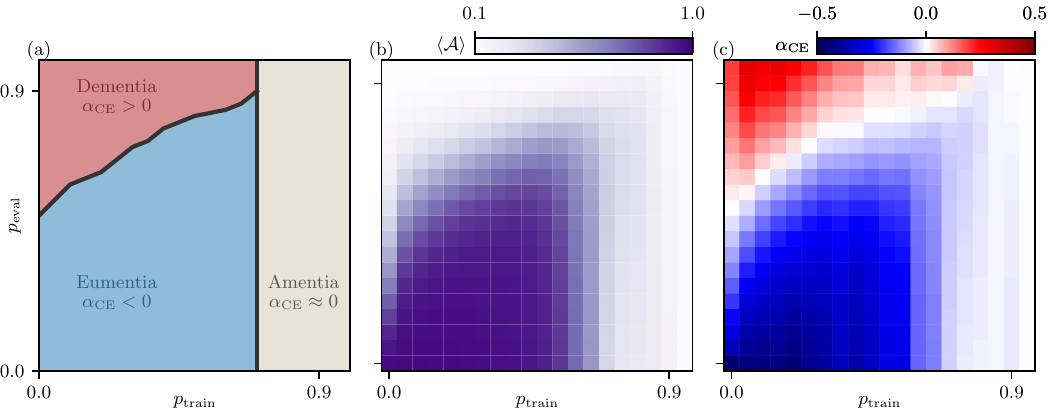}
    \caption{(a) Phase diagram as a function of the training dropout rate $p_{\text{train}}$ and evaluation dropout rate $p_{\text{eval}}$. The three phases eumentia, dementia, and amentia are characterized by the exponent of power-law decay in the cross-entropy loss as shown in panel (c); (b) Averaged accuracy $\expval{\mathcal{A}}$ for the training dataset size $N=60000$; (c) Exponent of cross-entropy loss $\alpha_{\text{CE}}$ defined in Eq.~\eqref{eq:CE} for the training dataset size $N=60000$, where the sign implies the three phases. 
    }
    \label{fig:CE_PD}
\end{figure*}

\para{Define the scaling behavior from the cross entropy and $\alpha_{\text{CE}}$}
Figure~\ref{fig:CE_PD}(a) presents the phase diagram in the $(p_{\text{train}}, p_{\text{eval}})$ plane. The averaged accuracy $\expval{\mathcal{A}}$ in Fig.~\ref{fig:CE_PD}(b) shows high accuracy concentrated at low dropout rates. The phase boundaries can be sharply determined by the exponent of the cross-entropy loss $\alpha_{\text{CE}}$ as shown in Fig.~\ref{fig:CE_PD}(c).
% \jp{JP: Please rearrange how you cite the figure so that Fig 2 (a),(b) cited in proper order} \HPcom{Done.}
Here, we find that the average test cross-entropy loss $\expval{\mathcal{L}_{\text{CE}}}$ exhibits a power-law scaling behavior as a function of the training dataset size $N$, consistent with previous studies on the scaling behavior of the test loss function~\cite{seung1992statistical, hestness2017deep, spigler2020asymptotic, kaplan2020scaling, rosenfeld2020constructive, bahri2020statistical,bordelon2020spectrum}
\begin{equation}\label{eq:CE}
    \expval{\mathcal{L}_{\text{CE}}}\sim N^{\alpha_{\text{CE}}}.
\end{equation}
We find that the exponent $\alpha_{\text{CE}}$ of the power-law scaling of the cross-entropy loss exhibits a negative value in the eumentia phase, implying that it can be trained since it keeps improving as we feed more data. 
However, in the dementia phase, we find a positive exponent $\alpha_{\text{CE}}$, implying that the dropout confuses the network and makes it worse as we feed more data. {At the pruning-induced transition between these two phases, we have $\alpha_{\text{CE}}\approx 0$}. 
In Sec.~\ref{sec:bkt}, we will show that near this transition the cross-entropy loss obeys a universal scaling function, consistent with a continuous phase transition.
In the amentia phase, we find $\alpha_{\text{CE}}$ is close to zero, which implies that the network cannot be trained effectively and the cross-entropy loss does not improve with more data. 

\para{Define three phases}
We summarize the three phases and their physical analogy as follows:
\begin{enumerate}
    \item \textbf{Eumentia} phase, $\alpha_{\text{CE}}<0$, appears at low training and evaluation dropout rates, where the network achieves high accuracy as shown in Fig.~\ref{fig:CE_PD}(b). This is analogous to a healthy brain with the ability to learn and generalize effectively. 
    \item \textbf{Dementia} phase, $\alpha_{\text{CE}}>0$, emerges at low training dropout but high evaluation dropout, where the network suffers from a significant drop in accuracy and a sharp increase in cross-entropy loss. This is reminiscent of a brain with severe cognitive impairment, where the network is trained well but fails to predict due to the high dropout rate during evaluation. 
    \item \textbf{Amentia} phase, $\alpha_{\text{CE}}\approx 0$, occurs at high training dropout, where the network fails to learn effectively regardless of the evaluation dropout rate, also resulting in low accuracy and high cross-entropy loss. This is analogous to a brain that is unable to learn at birth, where the network cannot be trained effectively due to the high dropout rate during training.
\end{enumerate}
In the next subsection, we will show the dependency of the cross-entropy loss as a function of the training dataset size $N$ more explicitly through the linecuts in Fig.~\ref{fig:CE_PD}.

\subsection{Exponents of the power-law scaling of the cross-entropy loss}
\begin{figure}[htbp]
    \centering
    \includegraphics[width=3.4in]{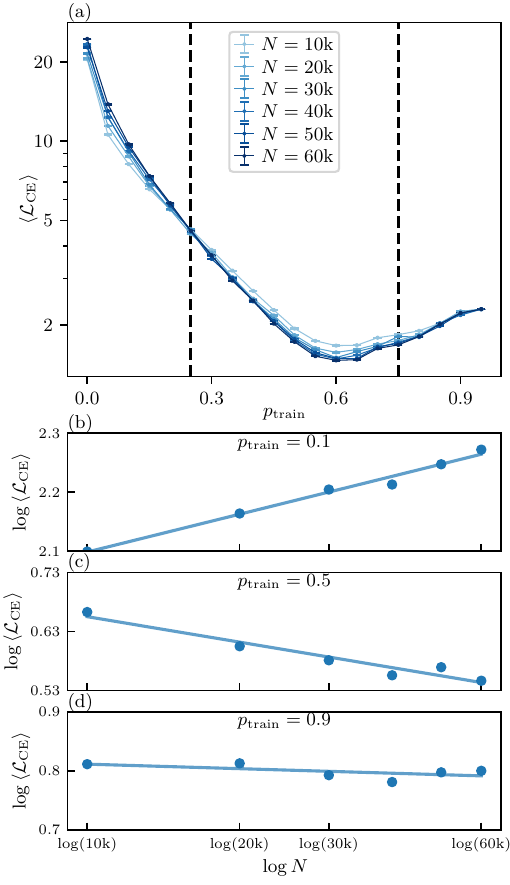}
    \caption{(a) Linecuts of the cross-entropy loss at a fixed evaluation dropout rate $p_{\text{eval}}=0.7$ as a function of the training dropout rate $p_{\text{train}}$ for different training dataset size $N$. 
    The vertical dashed line indicates the phase boundaries.
    (b-d) Power-law scaling of the averaged cross-entropy loss $\expval{\mathcal{L}_{\text{CE}}}$ as a function of the
    training dataset size $N$ at different training dropout rates $p_{\text{train}}$ in the dementia, eumentia, and amentia phases, respectively. The solid lines are power-law fits to the data in the log-log scale following Eq.~\eqref{eq:CE}.
    % \jp{JP: Too small too see. Please put (a) on top and (b) on bottom.}
    }
    \label{fig:CE_linecut_alpha}
\end{figure}

\para{Explain the three phases clearly}
To show the cross-entropy loss scaling more explicitly, we take the linecut of the cross-entropy loss landscape at a fixed $p_{\text{eval}}=0.7$ as a function of $p_{\text{train}}$ for different training sizes $N$ as shown in Fig.~\ref{fig:CE_linecut_alpha}(a).  
At a small training dropout rate $p_{\text{train}}\lessapprox 0.25$, we first observe the dementia phase where more data causes worse performance (high cross-entropy loss). 
As shown in Fig.~\ref{fig:CE_linecut_alpha}(b), the averaged cross-entropy loss $\expval{\mathcal{L}_{\text{CE}}}$ exhibits a power-law scaling behavior as a function of the training dataset $N$, following Eq.~\eqref{eq:CE}, with a positive exponent $\alpha_{\text{CE}}$.
At around $p_{\text{train}}\approx 0.25$, we find a crossing for different training dataset size $N$, which is a strong evidence of the phase transition, which we will study its critical behavior more carefully in the next section.
Beyond the crossing point, we find the transition into the eumentia phase, where the cross-entropy loss decreases algebraically as we feed more data, with a negative exponent $\alpha_{\text{CE}}$, following Eq.~\eqref{eq:CE}.
Finally, as we further increase $p_{\text{train}}\gtrapprox 0.7$, we find the transition into the amentia phase, with an almost constant cross-entropy loss for different training dataset size $N$.

% (Mention that for more details we can see)

% definition of $\alpha_\mathcal{A}$
\subsection{BKT-like phase transition from eumentia to dementia phase}\label{sec:bkt}
\begin{figure*}[htbp]
    \centering
    \includegraphics[width=7in]{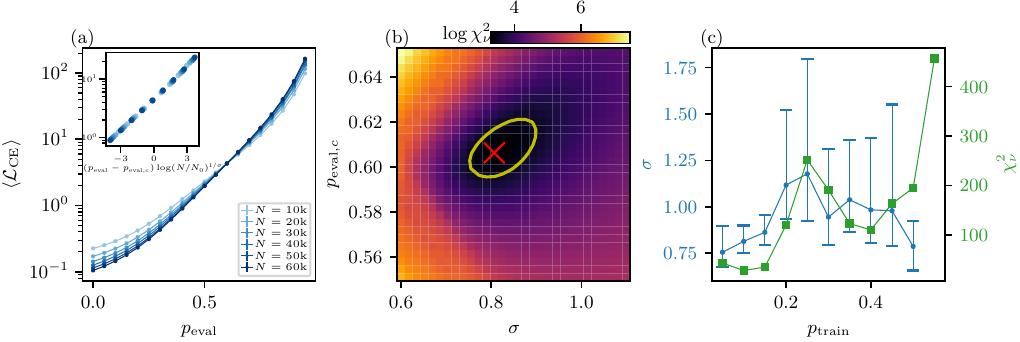}
    \caption{(a) Averaged cross-entropy loss as a function of evaluation dropout rate $p_{\text{eval}}$ at $p_{\text{train}}=0.1$ for different training dataset size $N$. Inset: the data collapse of the averaged cross-entropy loss $\expval{\mathcal{L}_{\text{CE}}}$ with the fitted value of $p_{\text{eval}, c}=0.60(2)$ and $\sigma=0.8(1)$. (b) Goodness of the data collapse $\chi_\nu^2$ as a function of the fitting parameter $\sigma$ and $p_{\text{eval}, c}$, where $\chi_{\nu,\min}^2$ is marked by the cross with value of 26.6, and the contour shows $1.3\chi_{\nu,\min}^2$. (c) Left axis: the fitted exponent $\sigma$ from the BKT-like transition ansatz in Eq.~\eqref{eq:bkt} as a function of the training dropout rate $p_{\text{train}}$;
    right axis: the corresponding goodness of the data collapse $\chi_\nu^2$.
    }
    \label{fig:CE_linecut_BKT}
\end{figure*}
\para{Motivate BKT transition from 2 points, we have large nu and power law decay, mention appendix, Explain BKT transition, define asatz clearly}
The crossing in the cross entropy data for various training sizes as we tune between the dementia and eumentia in Fig.~\ref{fig:CE_linecut_alpha}(a) is  strong evidence of a phase transition, which we find to be generic for all the transitions between the dementia and eumentia phase {(see Appendix~\ref{app:bkt})}. In Fig.~\ref{fig:CE_linecut_BKT}(a), we present the cross-entropy loss at a fixed training dropout rate $p_{\text{train}}=0.1$, showing the same crossing at around $p_{\text{eval}}\approx 0.6$.
A well-known feature of a trained neural network (i.e. the eumentia phase) is that the cross entropy decays as a power law in training size~\cite{seung1992statistical, hestness2017deep, spigler2020asymptotic, kaplan2020scaling, rosenfeld2020constructive, bahri2020statistical,bordelon2020spectrum}, Eq.~\eqref{eq:CE}, and does not decay exponentially. From a physics perspective of a phase of matter, the presence of power law decay in the ``ordered phase'' is indicative of QLRO instead of true long-range order. In the case of QLRO, the phase terminates at a BKT transition~\cite{berezinskii1971destruction,kosterlitz1973ordering,kosterlitz1974critical,kosterlitz2016kosterlitz} that describes the binding and unbinding of vortices with a correlation length that diverges like a stretched exponential instead of a simple power law. Motivated by these considerations, 
%Due to the power law decay of the cross entropy loss as a function of the training dataset size $N$, and the very large correlation length exponent $\nu$ extracted from the conventional algebraic divergence (i.e., assuming the correlation length diverges at critical point as $\xi\sim(p-p_c)^{-\nu}$, see Appendix~\ref{app:powerlaw} for more details), 
we conjecture that the transition between the eumentia and dementia phase is a BKT-like transition
%~\cite{berezinskii1971destruction,kosterlitz1973ordering,kosterlitz1974critical,kosterlitz2016kosterlitz} 
with the correlation length diverging as 
\begin{equation}
    \xi\sim \exp\left(\frac{a}{\abs{p-p_c}^{\sigma}}\right),
\end{equation} 
where $a$ and $\sigma$ are univeral parameters, and $p_c$ is the position of the critical point. 
We now generalize the conventional BKT scaling form, which has the exponent $\sigma=\frac{1}{2}$; here, we do not assume this value of $\sigma$ \textit{a priori} but assume only the stretched exponential form of the correlation length divergence {consistent with a generalized form of QLRO}. 
Therefore, we perform the finite-size scaling of the cross-entropy loss using the BKT-motivated ansatz,
\begin{equation}\label{eq:bkt}
    \expval{\mathcal{L}_{\text{CE}}} \sim f\left( (p_{\text{eval}}-p_{\text{eval}, c}) \left[ \log(N/N_0) \right]^{1/\sigma} \right),
\end{equation}
where $f$ is the universal scaling function, and $p_{\text{eval}, c}$ is the critical evaluation dropout rate at the transition between the eumentia and dementia phase, and $N_0$ is a microscopic scale, analogous to the lattice spacing that serves as the UV cutoff in the standard BKT transition. Since $N$ is of the order of $10^5$, and we find that the fitted result does not depend sensitively on  $N_0$, we fixed $N_0=1$ in the numerical fitting of the finite-size scaling to save one fitting parameter for better numerical stability.  

A contrasting scenario to the BKT-like scaling we present above is to assume the transition is a second order critical point with a power law divergence of the correlation length $\xi\sim(p-p_c)^{-\nu}$ that we explore in Appendix~\ref{app:powerlaw}. While we do find that the data also collapses following a second order transition scaling ansatz, we find that the quality of the collapse is, on average better using the BKT-like form and our estimate of the correlation length exponent is large $\nu \gtrapprox 8$. As the stretched exponential is consistent with $\nu\rightarrow \infty$, we find this analysis is consistent with our BKT scaling.

\para{Explain the plot of finite-size scaling}
We perform the finite-size scaling using the numerical package \texttt{fss}~\cite{pan2025fss}, and plot the loss function for the goodness of the data collapse $\chi_\nu^2$ \footnote{Note that the subscript $\nu$ here denotes the degrees of freedom used in the data collapse and should not be confused with the exponent of the algebraic divergence of the correlation length, as appearing in Eq. \ref{eq:powerlaw}}(see Appendix~\ref{app:bkt} for the definition) as a function of the fitting parameter $\sigma$ and $p_{\text{eval},c}$ in Fig.~\ref{fig:CE_linecut_BKT}(b), at a fixed $p_{\text{train}}=0.1$. We find a dip in $\chi_\nu^2$ at around $\sigma\approx 0.8$ and $p_{\text{eval},c}\approx 0.6$, with a closed contour enclosing the dip for $1.3\chi_{\nu,\min}^2$. This bounded contour implies a stable fitting result for the BKT-like transition. We present the fitting data collapse in the inset of Fig.~\ref{fig:CE_linecut_BKT}(a) with the fitting parameter of $\sigma= 0.8(1)$ and $p_{\text{eval},c}= 0.60(2)$. 

\para{Explain the final sweep}
Finally, we sweep the training dropout rate $p_{\text{train}}$ and study the entire phase transition from the eumentia to the dementia phase, and present the critical exponent $\sigma$ and the goodness of the data collapse $\chi_\nu^2$ as a function of $p_{\text{train}}$ in Fig.~\ref{fig:CE_linecut_BKT}(c).
The fitted exponent $\sigma$ ranges between approximately $0.8$ and $1.0$ (within the error bar) across the entire sweep of $p_{\text{train}}$ (blue curve in Fig.~\ref{fig:CE_linecut_BKT}(c)), consistently larger than the standard BKT value of $\sigma = \frac{1}{2}$.
The goodness of the data collapse $\chi_\nu^2$ (green curve in Fig.~\ref{fig:CE_linecut_BKT}(c)) remains small for small $p_{\text{train}}$, where the crossing of different training dataset sizes $N$ is well separated from the phase diagram boundary at $p_{\text{eval}} = 1$, yielding a clean and unambiguous data collapse. For large $p_{\text{train}}$, however, $\chi_\nu^2$ increases significantly, as the critical point $p_{\text{eval}, c}$ approaches the boundary $p_{\text{eval}} = 1$, where the cross-entropy loss curves for different $N$ become are sticked together (see Appendix~\ref{app:bkt} for details).

\subsection{Accuracy}
\begin{figure}[htbp]
    \centering
    \includegraphics[width=3.4in]{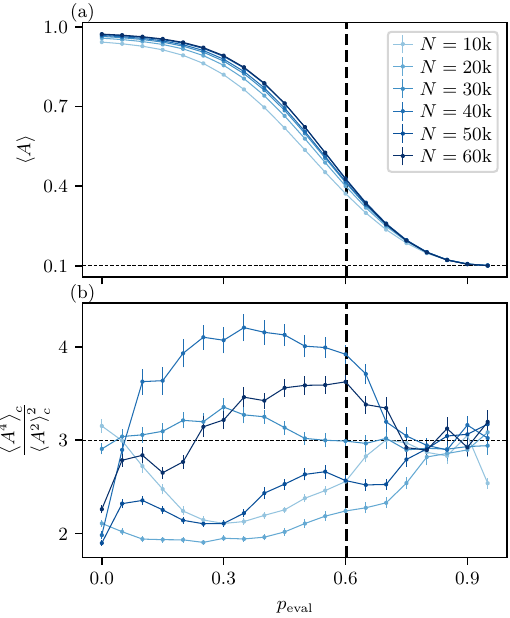}
    \caption{
        (a) Averaged accuracy $\expval{\mathcal{A}}$ at a fixed training dropout rate $p_{\text{train}}=0.1$ as a function of the evaluation dropout rate $p_{\text{eval}}$ for different training dataset size $N$. The vertical dashed line marks the critical point $p_{\text{eval}, c}$ obtained from Fig.~\ref{fig:CE_linecut_BKT} for all the panels. The horizontal dashed line marks the accuracy from random guessing. 
        (b) Binder cumulant (see Eq.~\ref{eq:binder} for the definition) of the same data in panel (a). The horizontal dashed line marks the expected value for a Gaussian variable.
    % \jp{JP: (1) Please put a vertical dashed line on each of these marking the transition? (2) Please use different dashed lines for 0.1 and 3 and explain them both here. (3) Is the amentia phase Gaussian too?  (4) Do we consider any statistics of the distribution of $\mathcal{L}_{CE}$ like we do here for accuracy? Also do we show these full distributions somewhere in each phase? It could be illuminating.}
    % \HPcom{
        % (1) Done; (2) Done 
    % (3) TODO: the current data only contains 20 train random seeds; 
    % (4) TODO: The variance and binder does not show clear feature for critical behavior though. 
    % }
    % \jp{JP: I would stack these two figures}
    % \HPcom{Fixed}
    % \jp{JP: I don't think the variance is adding much to the story here and we should remove it. It could be more interesting to show the accuracy vs $N$ in each phase as insets to (a).}
    }
    \label{fig:accuracy_stats}
\end{figure}

\para{Study the linecuts, along $p_{\text{train}}=0.1$ and then explain the mean of accuracy  }
Finally, we study the behavior of the averaged accuracy $\expval{\mathcal{A}}$ in Fig.~\ref{fig:accuracy_stats}(a) for $p_{\text{train}}=0.1$. We find that the accuracy maintains a high value around 1 in the eumentia phase, and drops to $0.1=1/10$ (which is the random guessing accuracy for 10 classes) in the dementia phase. The average accuracy keeps increasing in the eumentia phase as we feed more data, while it almost saturates in the dementia phase, qualitatively similar to the behavior of the cross-entropy loss.
However, the drop in the accuracy is not steep enough to allow us to make an accurate estimate of the location of the %show a 
clear critical point, unlike the cross-entropy loss that has a clear crossing.
% \jp{JP: Fix this in relation to where the transition is. I think the peak seems to be behind the transition right? I also think the accuracy becomes $N$ independent when you go across the transition as we will see in (a) when you mark the transition.}
% \HPcom{Indeed, now the vertical dashed line marked the $p_{\text{eval}, c}$ obtained from Fig.~\ref{fig:CE_linecut_BKT} }
% \para{Explain the variance and mention that the variance of accuracy also implies the phase transition but not quantitative enough to pinpoint the critical point}
% In Fig.~\ref{fig:accuracy_stats}(b), we present the variance of the accuracy $\expval{\mathcal{A}^2}-\expval{\mathcal{A}}^2$ at the same linecut of $p_{\text{train}}=0.1$. We find the peaks around $p_{\text{eval}}\approx 0.5$ for different training dataset size $N$, which is close to the criticality extracted from the cross-entropy loss in Fig.~\ref{fig:CE_linecut_BKT}(a). However, the peak position is not quantitatively accurate enough to pinpoint the criticality of the phase transition. 

\para{Binder cumulant showing 3 implies a Gaussian distribution }
Finally, we study the Binder cumulant of the accuracy, defined as 
\begin{equation}\label{eq:binder}
    U_4 = \frac{\expval{\mathcal{A}^4}_c}{\expval{\mathcal{A}^2}_c^2},
\end{equation}
where $\expval{\dots}_c$ is the central moment, i.e., $\expval{\mathcal{A}^n}_c = \expval{(\mathcal{A}-\expval{\mathcal{A}})^n}$. 
We find that the Binder cumulant $U_4$ approaches 3 in the dementia phase, implying a Gaussian distribution of the accuracy in the dementia phase {since it reduces to the kurtosis for a Gaussian distribution, which is 3.}
{The input $\bm{x}$ can be viewed as an $N$-body system interacting through the weight matrices $W$; in the dementia phase, dropout depletes the network so severely that the trained nonlinearities have no effective impact, reducing the system to a Gaussian (non-interacting) theory. This picture could be made quantifiable by examining how the effective interaction through the masked weight matrices depends on the dropout rate.}
% \jp{JP: Please quickly derive the Gaussian result and explain. } 
%\jp{It is rather interesting that we find the dementia phase is flowing to a Gaussian network, even though there are nonlinearities they can be viewed as irrelevant in this regime to the sparsity of their connections.}
% \HPcom{I feel like this part is a bit speculative}

\subsection{Generality of the phase diagram for different width and depth}
\begin{figure*}[ht]
    \centering
    \includegraphics[width=7in]{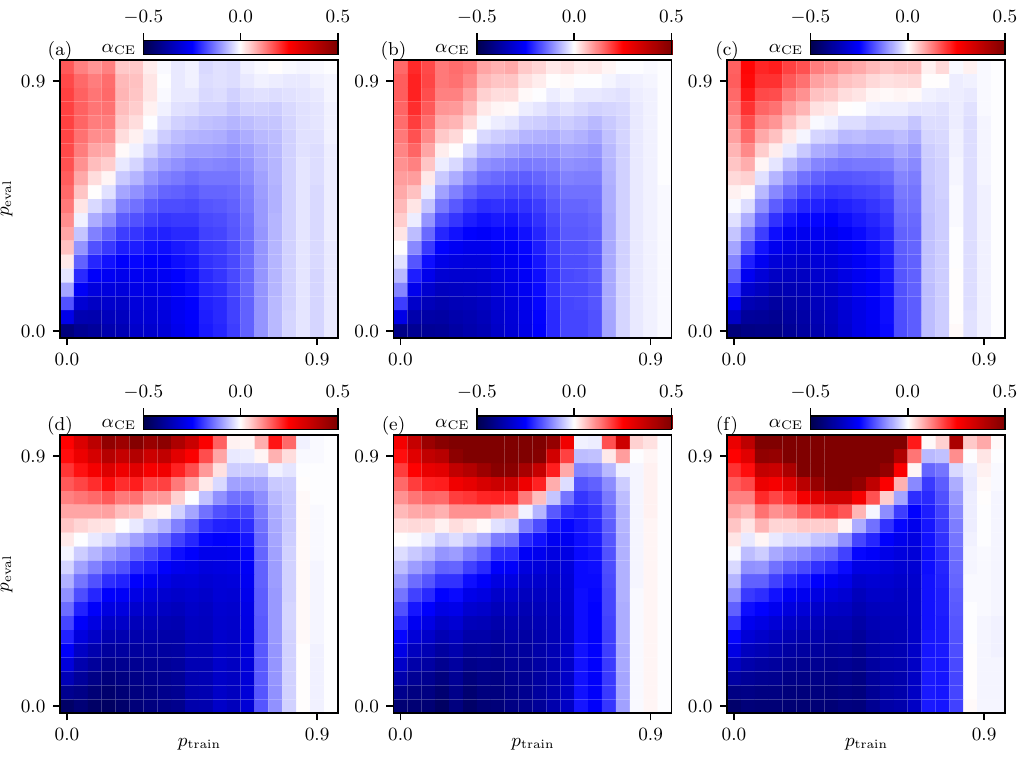}
    \caption{Exponent of cross-entropy loss $\alpha_{\text{CE}}$ for different width $n_h$ and depth $L$ for training size $N=60000$. (a-c) width $n_h=50$ and depth $L=2,3,4$; (d-f) depth $L=5$ and width $n_h=100, 150, 200$. }
    \label{fig:CE_w_l}
\end{figure*}
\para{We find it is generic, for the eumentia-amentia transition it is sharper cliff, for eumentia-dementia transition, is seems saturates, however, deeper NN are harder to train without using skip connection, which makes scaling to deeper NN hard}
Finally, we show that the three phases are generic for different widths $n_h$ and depths $L$ of the FCNN as shown in Fig.~\ref{fig:CE_w_l}. On the top row, we fixed the width $n_h=50$ and varied the depth $L=2,3,4$. On the bottom row, we fixed the depth $L=5$ and vary the width $n_h=100, 150, 200$. We find that although the phase boundaries vary for different widths and depths, the existence of the three phases is generic. Especially, the transition between the eumentia and amentia phase becomes sharper as we increase the width (as shown in Fig.~\ref{fig:CE_PD}(d-f)). 
For the transition between the eumentia and dementia phase, the phase boundary saturates as we increase the depth, as shown in Fig.~\ref{fig:CE_PD}(a-c).
However, since a deeper FCNN is harder to train without using skip connections, it is hard to scale to a deeper FCNN to further confirm the saturation of the phase boundary for the eumentia-dementia transition. 

\para{A line cut showing convergence}
\section{Conclusions and outlook}\label{eq:conclusion}
\para{Summary of the main results: (1) find three phases, can be characterized by the exponent of the power-law; (2) the transition between the eumentia and dementia phase is a BKT-like transition ; (3) the phase diagram is generic for different width and depth.}
In this paper, we study the phase diagram of an FCNN trained on MNIST by independently varying the training dropout rate $p_{\text{train}}$ and evaluation dropout rate $p_{\text{eval}}$. We identify three distinct phases---eumentia, dementia, and amentia---characterized by the sign of the exponent $\alpha_{\text{CE}}$ in the power-law scaling of the averaged test cross-entropy loss $\expval{\mathcal{L}_{\text{CE}}} $ with the training dataset size $N$: the eumentia phase ($\alpha_{\text{CE}} < 0$) where the network learns and improves with more data, the dementia phase ($\alpha_{\text{CE}} > 0$) where evaluation dropout degrades performance that worsens with more data, and the amentia phase ($\alpha_{\text{CE}} \approx 0$) where training dropout prevents the network from learning altogether even in the training stage. 
We further show that the transition between the eumentia and dementia phases suggests a BKT-like transition, with finite-size scaling analysis yielding a critical exponent $\sigma \approx 0.8$--$1.0$, consistently larger than the standard BKT value of $\sigma = 1/2$. 
Finally, we demonstrate that the three-phase structure is generic across different network widths $n_h$ and depths $L$ of the FCNN.

\para{Future: (1)scaling for width and depth, (2) change the architecture, change the dataset, activation function, and optimizer; (3) "temporal" dynamics ; (4) for the stat mech models; (5) correlation functions}
An intriguing direction is to explore the connection between the dropout-induced phases and synaptic pruning in biological neural networks, where the selective elimination of synapses during development is essential for brain maturation~\cite{huttenlocher1979synaptic,chechik1998synaptic, navlakha2015decreasingrate}.
It is also natural to investigate whether the three-phase structure and the BKT-like transition persist in other network architectures and learning tasks.

\section*{Acknowledgement}
We thank Antoine Georges, David Huse, Sankar Das Sarma, Jacob Taylor, Agnes Valenti, and Jiabin Yu for fruitful discussions.
This work is partially supported by US-ONR grant No.~N00014-23-1-2357 (H.P. and J.H.P.). This work was partially conceived at the Kavli Institute of Theoretical Physics which is supported in part by the National Science Foundation under Grants No.~NSF PHY-1748958 and PHY-2309135 (J.H.P.)

\bibliography{references, Paper_RNN}

\begin{thebibliography}{43}%
\makeatletter
\providecommand \@ifxundefined [1]{%
 \@ifx{#1\undefined}
}%
\providecommand \@ifnum [1]{%
 \ifnum #1\expandafter \@firstoftwo
 \else \expandafter \@secondoftwo
 \fi
}%
\providecommand \@ifx [1]{%
 \ifx #1\expandafter \@firstoftwo
 \else \expandafter \@secondoftwo
 \fi
}%
\providecommand \natexlab [1]{#1}%
\providecommand \enquote  [1]{``#1''}%
\providecommand \bibnamefont  [1]{#1}%
\providecommand \bibfnamefont [1]{#1}%
\providecommand \citenamefont [1]{#1}%
\providecommand \href@noop [0]{\@secondoftwo}%
\providecommand \href [0]{\begingroup \@sanitize@url \@href}%
\providecommand \@href[1]{\@@startlink{#1}\@@href}%
\providecommand \@@href[1]{\endgroup#1\@@endlink}%
\providecommand \@sanitize@url [0]{\catcode `\\12\catcode `\$12\catcode `\&12\catcode `\#12\catcode `\^12\catcode `\_12\catcode `\%12\relax}%
\providecommand \@@startlink[1]{}%
\providecommand \@@endlink[0]{}%
\providecommand \url  [0]{\begingroup\@sanitize@url \@url }%
\providecommand \@url [1]{\endgroup\@href {#1}{\urlprefix }}%
\providecommand \urlprefix  [0]{URL }%
\providecommand \Eprint [0]{\href }%
\providecommand \doibase [0]{https://doi.org/}%
\providecommand \selectlanguage [0]{\@gobble}%
\providecommand \bibinfo  [0]{\@secondoftwo}%
\providecommand \bibfield  [0]{\@secondoftwo}%
\providecommand \translation [1]{[#1]}%
\providecommand \BibitemOpen [0]{}%
\providecommand \bibitemStop [0]{}%
\providecommand \bibitemNoStop [0]{.\EOS\space}%
\providecommand \EOS [0]{\spacefactor3000\relax}%
\providecommand \BibitemShut  [1]{\csname bibitem#1\endcsname}%
\let\auto@bib@innerbib\@empty
%</preamble>
\bibitem [{\citenamefont {Brown}\ \emph {et~al.}(2020)\citenamefont {Brown}, \citenamefont {Mann}, \citenamefont {Ryder}, \citenamefont {Subbiah}, \citenamefont {Kaplan}, \citenamefont {Dhariwal}, \citenamefont {Neelakantan}, \citenamefont {Shyam}, \citenamefont {Sastry}, \citenamefont {Askell}, \citenamefont {Agarwal}, \citenamefont {{Herbert-Voss}}, \citenamefont {Krueger}, \citenamefont {Henighan}, \citenamefont {Child}, \citenamefont {Ramesh}, \citenamefont {Ziegler}, \citenamefont {Wu}, \citenamefont {Winter}, \citenamefont {Hesse}, \citenamefont {Chen}, \citenamefont {Sigler}, \citenamefont {Litwin}, \citenamefont {Gray}, \citenamefont {Chess}, \citenamefont {Clark}, \citenamefont {Berner}, \citenamefont {McCandlish}, \citenamefont {Radford}, \citenamefont {Sutskever},\ and\ \citenamefont {Amodei}}]{brown2020language}%
  \BibitemOpen
  \bibfield  {author} {\bibinfo {author} {\bibfnamefont {T.}~\bibnamefont {Brown}}, \bibinfo {author} {\bibfnamefont {B.}~\bibnamefont {Mann}}, \bibinfo {author} {\bibfnamefont {N.}~\bibnamefont {Ryder}}, \bibinfo {author} {\bibfnamefont {M.}~\bibnamefont {Subbiah}}, \bibinfo {author} {\bibfnamefont {J.~D.}\ \bibnamefont {Kaplan}}, \bibinfo {author} {\bibfnamefont {P.}~\bibnamefont {Dhariwal}}, \bibinfo {author} {\bibfnamefont {A.}~\bibnamefont {Neelakantan}}, \bibinfo {author} {\bibfnamefont {P.}~\bibnamefont {Shyam}}, \bibinfo {author} {\bibfnamefont {G.}~\bibnamefont {Sastry}}, \bibinfo {author} {\bibfnamefont {A.}~\bibnamefont {Askell}}, \bibinfo {author} {\bibfnamefont {S.}~\bibnamefont {Agarwal}}, \bibinfo {author} {\bibfnamefont {A.}~\bibnamefont {{Herbert-Voss}}}, \bibinfo {author} {\bibfnamefont {G.}~\bibnamefont {Krueger}}, \bibinfo {author} {\bibfnamefont {T.}~\bibnamefont {Henighan}}, \bibinfo {author} {\bibfnamefont {R.}~\bibnamefont {Child}}, \bibinfo {author} {\bibfnamefont {A.}~\bibnamefont {Ramesh}}, \bibinfo {author} {\bibfnamefont {D.}~\bibnamefont {Ziegler}}, \bibinfo {author} {\bibfnamefont {J.}~\bibnamefont {Wu}}, \bibinfo {author} {\bibfnamefont {C.}~\bibnamefont {Winter}}, \bibinfo {author} {\bibfnamefont {C.}~\bibnamefont {Hesse}}, \bibinfo {author} {\bibfnamefont {M.}~\bibnamefont {Chen}}, \bibinfo {author} {\bibfnamefont {E.}~\bibnamefont {Sigler}}, \bibinfo {author} {\bibfnamefont {M.}~\bibnamefont {Litwin}}, \bibinfo {author} {\bibfnamefont {S.}~\bibnamefont {Gray}}, \bibinfo {author} {\bibfnamefont {B.}~\bibnamefont {Chess}}, \bibinfo {author} {\bibfnamefont {J.}~\bibnamefont {Clark}}, \bibinfo {author} {\bibfnamefont {C.}~\bibnamefont {Berner}}, \bibinfo {author} {\bibfnamefont {S.}~\bibnamefont {McCandlish}}, \bibinfo {author} {\bibfnamefont {A.}~\bibnamefont {Radford}}, \bibinfo {author} {\bibfnamefont {I.}~\bibnamefont {Sutskever}},\ and\ \bibinfo {author} {\bibfnamefont {D.}~\bibnamefont {Amodei}},\ }\bibfield  {title} {\bibinfo {title} {Language {{Models}} are {{Few-Shot Learners}}},\ }in\ \href {https://papers.nips.cc/paper/2020/hash/1457c0d6bfcb4967418bfb8ac142f64a-Abstract.html} {\emph {\bibinfo {booktitle} {Advances in {{Neural Information Processing Systems}}}}},\ Vol.~\bibinfo {volume} {33}\ (\bibinfo  {publisher} {Curran Associates, Inc.},\ \bibinfo {year} {2020})\ pp.\ \bibinfo {pages} {1877--1901}\BibitemShut {NoStop}%
\bibitem [{\citenamefont {Touvron}\ \emph {et~al.}(2023)\citenamefont {Touvron}, \citenamefont {Lavril}, \citenamefont {Izacard}, \citenamefont {Martinet}, \citenamefont {Lachaux}, \citenamefont {Lacroix}, \citenamefont {Rozi{\`e}re}, \citenamefont {Goyal}, \citenamefont {Hambro}, \citenamefont {Azhar}, \citenamefont {Rodriguez}, \citenamefont {Joulin}, \citenamefont {Grave},\ and\ \citenamefont {Lample}}]{touvron2023llama}%
  \BibitemOpen
  \bibfield  {author} {\bibinfo {author} {\bibfnamefont {H.}~\bibnamefont {Touvron}}, \bibinfo {author} {\bibfnamefont {T.}~\bibnamefont {Lavril}}, \bibinfo {author} {\bibfnamefont {G.}~\bibnamefont {Izacard}}, \bibinfo {author} {\bibfnamefont {X.}~\bibnamefont {Martinet}}, \bibinfo {author} {\bibfnamefont {M.-A.}\ \bibnamefont {Lachaux}}, \bibinfo {author} {\bibfnamefont {T.}~\bibnamefont {Lacroix}}, \bibinfo {author} {\bibfnamefont {B.}~\bibnamefont {Rozi{\`e}re}}, \bibinfo {author} {\bibfnamefont {N.}~\bibnamefont {Goyal}}, \bibinfo {author} {\bibfnamefont {E.}~\bibnamefont {Hambro}}, \bibinfo {author} {\bibfnamefont {F.}~\bibnamefont {Azhar}}, \bibinfo {author} {\bibfnamefont {A.}~\bibnamefont {Rodriguez}}, \bibinfo {author} {\bibfnamefont {A.}~\bibnamefont {Joulin}}, \bibinfo {author} {\bibfnamefont {E.}~\bibnamefont {Grave}},\ and\ \bibinfo {author} {\bibfnamefont {G.}~\bibnamefont {Lample}},\ }\href {http://arxiv.org/abs/2302.13971} {\bibinfo {title} {{{LLaMA}}: {{Open}} and {{Efficient Foundation Language Models}}}} (\bibinfo {year} {2023})\BibitemShut {NoStop}%
\bibitem [{\citenamefont {Chowdhery}\ \emph {et~al.}(2023)\citenamefont {Chowdhery}, \citenamefont {Narang}, \citenamefont {Devlin}, \citenamefont {Bosma}, \citenamefont {Mishra}, \citenamefont {Roberts}, \citenamefont {Barham}, \citenamefont {Chung}, \citenamefont {Sutton}, \citenamefont {Gehrmann}, \citenamefont {Schuh}, \citenamefont {Shi}, \citenamefont {Tsvyashchenko}, \citenamefont {Maynez}, \citenamefont {Rao}, \citenamefont {Barnes}, \citenamefont {Tay}, \citenamefont {Shazeer}, \citenamefont {Prabhakaran}, \citenamefont {Reif}, \citenamefont {Du}, \citenamefont {Hutchinson}, \citenamefont {Pope}, \citenamefont {Bradbury}, \citenamefont {Austin}, \citenamefont {Isard}, \citenamefont {{Gur-Ari}}, \citenamefont {Yin}, \citenamefont {Duke}, \citenamefont {Levskaya}, \citenamefont {Ghemawat}, \citenamefont {Dev}, \citenamefont {Michalewski}, \citenamefont {Garcia}, \citenamefont {Misra}, \citenamefont {Robinson}, \citenamefont {Fedus}, \citenamefont {Zhou}, \citenamefont {Ippolito}, \citenamefont {Luan}, \citenamefont {Lim}, \citenamefont {Zoph}, \citenamefont {Spiridonov}, \citenamefont {Sepassi}, \citenamefont {Dohan}, \citenamefont {Agrawal}, \citenamefont {Omernick}, \citenamefont {Dai}, \citenamefont {Pillai}, \citenamefont {Pellat}, \citenamefont {Lewkowycz}, \citenamefont {Moreira}, \citenamefont {Child}, \citenamefont {Polozov}, \citenamefont {Lee}, \citenamefont {Zhou}, \citenamefont {Wang}, \citenamefont {Saeta}, \citenamefont {Diaz}, \citenamefont {Firat}, \citenamefont {Catasta}, \citenamefont {Wei}, \citenamefont {{Meier-Hellstern}}, \citenamefont {Eck}, \citenamefont {Dean}, \citenamefont {Petrov},\ and\ \citenamefont {Fiedel}}]{chowdhery2023palm}%
  \BibitemOpen
  \bibfield  {author} {\bibinfo {author} {\bibfnamefont {A.}~\bibnamefont {Chowdhery}}, \bibinfo {author} {\bibfnamefont {S.}~\bibnamefont {Narang}}, \bibinfo {author} {\bibfnamefont {J.}~\bibnamefont {Devlin}}, \bibinfo {author} {\bibfnamefont {M.}~\bibnamefont {Bosma}}, \bibinfo {author} {\bibfnamefont {G.}~\bibnamefont {Mishra}}, \bibinfo {author} {\bibfnamefont {A.}~\bibnamefont {Roberts}}, \bibinfo {author} {\bibfnamefont {P.}~\bibnamefont {Barham}}, \bibinfo {author} {\bibfnamefont {H.~W.}\ \bibnamefont {Chung}}, \bibinfo {author} {\bibfnamefont {C.}~\bibnamefont {Sutton}}, \bibinfo {author} {\bibfnamefont {S.}~\bibnamefont {Gehrmann}}, \bibinfo {author} {\bibfnamefont {P.}~\bibnamefont {Schuh}}, \bibinfo {author} {\bibfnamefont {K.}~\bibnamefont {Shi}}, \bibinfo {author} {\bibfnamefont {S.}~\bibnamefont {Tsvyashchenko}}, \bibinfo {author} {\bibfnamefont {J.}~\bibnamefont {Maynez}}, \bibinfo {author} {\bibfnamefont {A.}~\bibnamefont {Rao}}, \bibinfo {author} {\bibfnamefont {P.}~\bibnamefont {Barnes}}, \bibinfo {author} {\bibfnamefont {Y.}~\bibnamefont {Tay}}, \bibinfo {author} {\bibfnamefont {N.}~\bibnamefont {Shazeer}}, \bibinfo {author} {\bibfnamefont {V.}~\bibnamefont {Prabhakaran}}, \bibinfo {author} {\bibfnamefont {E.}~\bibnamefont {Reif}}, \bibinfo {author} {\bibfnamefont {N.}~\bibnamefont {Du}}, \bibinfo {author} {\bibfnamefont {B.}~\bibnamefont {Hutchinson}}, \bibinfo {author} {\bibfnamefont {R.}~\bibnamefont {Pope}}, \bibinfo {author} {\bibfnamefont {J.}~\bibnamefont {Bradbury}}, \bibinfo {author} {\bibfnamefont {J.}~\bibnamefont {Austin}}, \bibinfo {author} {\bibfnamefont {M.}~\bibnamefont {Isard}}, \bibinfo {author} {\bibfnamefont {G.}~\bibnamefont {{Gur-Ari}}}, \bibinfo {author} {\bibfnamefont {P.}~\bibnamefont {Yin}}, \bibinfo {author} {\bibfnamefont {T.}~\bibnamefont {Duke}}, \bibinfo {author} {\bibfnamefont {A.}~\bibnamefont {Levskaya}}, \bibinfo {author} {\bibfnamefont {S.}~\bibnamefont {Ghemawat}}, \bibinfo {author} {\bibfnamefont {S.}~\bibnamefont {Dev}}, \bibinfo {author} {\bibfnamefont {H.}~\bibnamefont {Michalewski}}, \bibinfo {author} {\bibfnamefont {X.}~\bibnamefont {Garcia}}, \bibinfo {author} {\bibfnamefont {V.}~\bibnamefont {Misra}}, \bibinfo {author} {\bibfnamefont {K.}~\bibnamefont {Robinson}}, \bibinfo {author} {\bibfnamefont {L.}~\bibnamefont {Fedus}}, \bibinfo {author} {\bibfnamefont {D.}~\bibnamefont {Zhou}}, \bibinfo {author} {\bibfnamefont {D.}~\bibnamefont {Ippolito}}, \bibinfo {author} {\bibfnamefont {D.}~\bibnamefont {Luan}}, \bibinfo {author} {\bibfnamefont {H.}~\bibnamefont {Lim}}, \bibinfo {author} {\bibfnamefont {B.}~\bibnamefont {Zoph}}, \bibinfo {author} {\bibfnamefont {A.}~\bibnamefont {Spiridonov}}, \bibinfo {author} {\bibfnamefont {R.}~\bibnamefont {Sepassi}}, \bibinfo {author} {\bibfnamefont {D.}~\bibnamefont {Dohan}}, \bibinfo {author} {\bibfnamefont {S.}~\bibnamefont {Agrawal}}, \bibinfo {author} {\bibfnamefont {M.}~\bibnamefont {Omernick}}, \bibinfo {author} {\bibfnamefont {A.~M.}\ \bibnamefont {Dai}}, \bibinfo {author} {\bibfnamefont {T.~S.}\ \bibnamefont {Pillai}}, \bibinfo {author} {\bibfnamefont {M.}~\bibnamefont {Pellat}}, \bibinfo {author} {\bibfnamefont {A.}~\bibnamefont {Lewkowycz}}, \bibinfo {author} {\bibfnamefont {E.}~\bibnamefont {Moreira}}, \bibinfo {author} {\bibfnamefont {R.}~\bibnamefont {Child}}, \bibinfo {author} {\bibfnamefont {O.}~\bibnamefont {Polozov}}, \bibinfo {author} {\bibfnamefont {K.}~\bibnamefont {Lee}}, \bibinfo {author} {\bibfnamefont {Z.}~\bibnamefont {Zhou}}, \bibinfo {author} {\bibfnamefont {X.}~\bibnamefont {Wang}}, \bibinfo {author} {\bibfnamefont {B.}~\bibnamefont {Saeta}}, \bibinfo {author} {\bibfnamefont {M.}~\bibnamefont {Diaz}}, \bibinfo {author} {\bibfnamefont {O.}~\bibnamefont {Firat}}, \bibinfo {author} {\bibfnamefont {M.}~\bibnamefont {Catasta}}, \bibinfo {author} {\bibfnamefont {J.}~\bibnamefont {Wei}}, \bibinfo {author} {\bibfnamefont {K.}~\bibnamefont {{Meier-Hellstern}}}, \bibinfo {author} {\bibfnamefont {D.}~\bibnamefont {Eck}}, \bibinfo {author} {\bibfnamefont {J.}~\bibnamefont {Dean}}, \bibinfo {author} {\bibfnamefont {S.}~\bibnamefont {Petrov}},\ and\ \bibinfo {author} {\bibfnamefont {N.}~\bibnamefont {Fiedel}},\ }\bibfield  {title} {\bibinfo {title} {{{PaLM}}: {{Scaling Language Modeling}} with {{Pathways}}},\ }\href {http://jmlr.org/papers/v24/22-1144.html} {\bibfield  {journal} {\bibinfo  {journal} {Journal of Machine Learning Research}\ }\textbf {\bibinfo {volume} {24}},\ \bibinfo {pages} {1} (\bibinfo {year} {2023})}\BibitemShut {NoStop}%
\bibitem [{\citenamefont {Gromov}\ \emph {et~al.}(2024)\citenamefont {Gromov}, \citenamefont {Tirumala}, \citenamefont {Shapourian}, \citenamefont {Glorioso},\ and\ \citenamefont {Roberts}}]{gromov2024unreasonable}%
  \BibitemOpen
  \bibfield  {author} {\bibinfo {author} {\bibfnamefont {A.}~\bibnamefont {Gromov}}, \bibinfo {author} {\bibfnamefont {K.}~\bibnamefont {Tirumala}}, \bibinfo {author} {\bibfnamefont {H.}~\bibnamefont {Shapourian}}, \bibinfo {author} {\bibfnamefont {P.}~\bibnamefont {Glorioso}},\ and\ \bibinfo {author} {\bibfnamefont {D.}~\bibnamefont {Roberts}},\ }\bibfield  {title} {\bibinfo {title} {The {{Unreasonable Ineffectiveness}} of the {{Deeper Layers}}},\ }in\ \href {https://openreview.net/forum?id=ngmEcEer8a} {\emph {\bibinfo {booktitle} {The {{Thirteenth International Conference}} on {{Learning Representations}}}}}\ (\bibinfo {year} {2024})\BibitemShut {NoStop}%
\bibitem [{\citenamefont {Han}\ \emph {et~al.}(2015)\citenamefont {Han}, \citenamefont {Pool}, \citenamefont {Tran},\ and\ \citenamefont {Dally}}]{han2015learning}%
  \BibitemOpen
  \bibfield  {author} {\bibinfo {author} {\bibfnamefont {S.}~\bibnamefont {Han}}, \bibinfo {author} {\bibfnamefont {J.}~\bibnamefont {Pool}}, \bibinfo {author} {\bibfnamefont {J.}~\bibnamefont {Tran}},\ and\ \bibinfo {author} {\bibfnamefont {W.~J.}\ \bibnamefont {Dally}},\ }\bibfield  {title} {\bibinfo {title} {Learning both weights and connections for efficient neural networks},\ }in\ \href@noop {} {\emph {\bibinfo {booktitle} {Proceedings of the 29th {{International Conference}} on {{Neural Information Processing Systems}} - {{Volume}} 1}}},\ \bibinfo {series} {{{NIPS}}'15}, Vol.~\bibinfo {volume} {1}\ (\bibinfo  {publisher} {MIT Press},\ \bibinfo {address} {Cambridge, MA, USA},\ \bibinfo {year} {2015})\ pp.\ \bibinfo {pages} {1135--1143}\BibitemShut {NoStop}%
\bibitem [{\citenamefont {Li}\ \emph {et~al.}(2017)\citenamefont {Li}, \citenamefont {Kadav}, \citenamefont {Durdanovic}, \citenamefont {Samet},\ and\ \citenamefont {Graf}}]{li2017pruning}%
  \BibitemOpen
  \bibfield  {author} {\bibinfo {author} {\bibfnamefont {H.}~\bibnamefont {Li}}, \bibinfo {author} {\bibfnamefont {A.}~\bibnamefont {Kadav}}, \bibinfo {author} {\bibfnamefont {I.}~\bibnamefont {Durdanovic}}, \bibinfo {author} {\bibfnamefont {H.}~\bibnamefont {Samet}},\ and\ \bibinfo {author} {\bibfnamefont {H.~P.}\ \bibnamefont {Graf}},\ }\bibfield  {title} {\bibinfo {title} {Pruning {{Filters}} for {{Efficient ConvNets}}},\ }in\ \href {https://openreview.net/forum?id=rJqFGTslg} {\emph {\bibinfo {booktitle} {International {{Conference}} on {{Learning Representations}}}}}\ (\bibinfo {year} {2017})\BibitemShut {NoStop}%
\bibitem [{\citenamefont {Men}\ \emph {et~al.}(2025)\citenamefont {Men}, \citenamefont {Xu}, \citenamefont {Zhang}, \citenamefont {Yuan}, \citenamefont {Wang}, \citenamefont {Lin}, \citenamefont {Lu}, \citenamefont {Han},\ and\ \citenamefont {Chen}}]{men2025shortgpt}%
  \BibitemOpen
  \bibfield  {author} {\bibinfo {author} {\bibfnamefont {X.}~\bibnamefont {Men}}, \bibinfo {author} {\bibfnamefont {M.}~\bibnamefont {Xu}}, \bibinfo {author} {\bibfnamefont {Q.}~\bibnamefont {Zhang}}, \bibinfo {author} {\bibfnamefont {Q.}~\bibnamefont {Yuan}}, \bibinfo {author} {\bibfnamefont {B.}~\bibnamefont {Wang}}, \bibinfo {author} {\bibfnamefont {H.}~\bibnamefont {Lin}}, \bibinfo {author} {\bibfnamefont {Y.}~\bibnamefont {Lu}}, \bibinfo {author} {\bibfnamefont {X.}~\bibnamefont {Han}},\ and\ \bibinfo {author} {\bibfnamefont {W.}~\bibnamefont {Chen}},\ }\bibfield  {title} {\bibinfo {title} {{{ShortGPT}}: {{Layers}} in {{Large Language Models}} are {{More Redundant Than You Expect}}},\ }in\ \href {https://aclanthology.org/2025.findings-acl.1035/} {\emph {\bibinfo {booktitle} {Findings of the {{Association}} for {{Computational Linguistics}}: {{ACL}} 2025}}},\ \bibinfo {editor} {edited by\ \bibinfo {editor} {\bibfnamefont {W.}~\bibnamefont {Che}}, \bibinfo {editor} {\bibfnamefont {J.}~\bibnamefont {Nabende}}, \bibinfo {editor} {\bibfnamefont {E.}~\bibnamefont {Shutova}},\ and\ \bibinfo {editor} {\bibfnamefont {M.~T.}\ \bibnamefont {Pilehvar}}}\ (\bibinfo  {publisher} {Association for Computational Linguistics},\ \bibinfo {address} {Vienna, Austria},\ \bibinfo {year} {2025})\ pp.\ \bibinfo {pages} {20192--20204}\BibitemShut {NoStop}%
\bibitem [{\citenamefont {Hoefler}\ \emph {et~al.}(2021)\citenamefont {Hoefler}, \citenamefont {Alistarh}, \citenamefont {{Ben-Nun}}, \citenamefont {Dryden},\ and\ \citenamefont {Peste}}]{hoefler2021sparsity}%
  \BibitemOpen
  \bibfield  {author} {\bibinfo {author} {\bibfnamefont {T.}~\bibnamefont {Hoefler}}, \bibinfo {author} {\bibfnamefont {D.}~\bibnamefont {Alistarh}}, \bibinfo {author} {\bibfnamefont {T.}~\bibnamefont {{Ben-Nun}}}, \bibinfo {author} {\bibfnamefont {N.}~\bibnamefont {Dryden}},\ and\ \bibinfo {author} {\bibfnamefont {A.}~\bibnamefont {Peste}},\ }\bibfield  {title} {\bibinfo {title} {Sparsity in deep learning: Pruning and growth for efficient inference and training in neural networks},\ }\href {https://dl.acm.org/doi/10.5555/3546258.3546499} {\bibfield  {journal} {\bibinfo  {journal} {J. Mach. Learn. Res.}\ }\textbf {\bibinfo {volume} {22}},\ \bibinfo {pages} {241:10882} (\bibinfo {year} {2021})}\BibitemShut {NoStop}%
\bibitem [{\citenamefont {LeCun}\ \emph {et~al.}(1989)\citenamefont {LeCun}, \citenamefont {Denker},\ and\ \citenamefont {Solla}}]{lecun1989optimal}%
  \BibitemOpen
  \bibfield  {author} {\bibinfo {author} {\bibfnamefont {Y.}~\bibnamefont {LeCun}}, \bibinfo {author} {\bibfnamefont {J.}~\bibnamefont {Denker}},\ and\ \bibinfo {author} {\bibfnamefont {S.}~\bibnamefont {Solla}},\ }\bibfield  {title} {\bibinfo {title} {Optimal {{Brain Damage}}},\ }in\ \href {https://proceedings.neurips.cc/paper/1989/hash/6c9882bbac1c7093bd25041881277658-Abstract.html} {\emph {\bibinfo {booktitle} {Advances in {{Neural Information Processing Systems}}}}},\ Vol.~\bibinfo {volume} {2}\ (\bibinfo  {publisher} {Morgan-Kaufmann},\ \bibinfo {year} {1989})\BibitemShut {NoStop}%
\bibitem [{\citenamefont {Hassibi}\ \emph {et~al.}(1993)\citenamefont {Hassibi}, \citenamefont {Stork},\ and\ \citenamefont {Wolff}}]{hassibi1993optimal}%
  \BibitemOpen
  \bibfield  {author} {\bibinfo {author} {\bibfnamefont {B.}~\bibnamefont {Hassibi}}, \bibinfo {author} {\bibfnamefont {D.}~\bibnamefont {Stork}},\ and\ \bibinfo {author} {\bibfnamefont {G.}~\bibnamefont {Wolff}},\ }\bibfield  {title} {\bibinfo {title} {Optimal {{Brain Surgeon}} and general network pruning},\ }in\ \href {https://ieeexplore.ieee.org/document/298572} {\emph {\bibinfo {booktitle} {{{IEEE International Conference}} on {{Neural Networks}}}}}\ (\bibinfo {year} {1993})\ pp.\ \bibinfo {pages} {293--299 vol.1}\BibitemShut {NoStop}%
\bibitem [{\citenamefont {Wen}\ \emph {et~al.}(2016)\citenamefont {Wen}, \citenamefont {Wu}, \citenamefont {Wang}, \citenamefont {Chen},\ and\ \citenamefont {Li}}]{wen2016learning}%
  \BibitemOpen
  \bibfield  {author} {\bibinfo {author} {\bibfnamefont {W.}~\bibnamefont {Wen}}, \bibinfo {author} {\bibfnamefont {C.}~\bibnamefont {Wu}}, \bibinfo {author} {\bibfnamefont {Y.}~\bibnamefont {Wang}}, \bibinfo {author} {\bibfnamefont {Y.}~\bibnamefont {Chen}},\ and\ \bibinfo {author} {\bibfnamefont {H.}~\bibnamefont {Li}},\ }\bibfield  {title} {\bibinfo {title} {Learning structured sparsity in deep neural networks},\ }in\ \href {https://dl.acm.org/doi/10.5555/3157096.3157329} {\emph {\bibinfo {booktitle} {Proceedings of the 30th {{International Conference}} on {{Neural Information Processing Systems}}}}},\ \bibinfo {series and number} {{{NIPS}}'16}\ (\bibinfo  {publisher} {Curran Associates Inc.},\ \bibinfo {address} {Red Hook, NY, USA},\ \bibinfo {year} {2016})\ pp.\ \bibinfo {pages} {2082--2090}\BibitemShut {NoStop}%
\bibitem [{\citenamefont {Srivastava}\ \emph {et~al.}(2014)\citenamefont {Srivastava}, \citenamefont {Hinton}, \citenamefont {Krizhevsky}, \citenamefont {Sutskever},\ and\ \citenamefont {Salakhutdinov}}]{srivastava2014dropout}%
  \BibitemOpen
  \bibfield  {author} {\bibinfo {author} {\bibfnamefont {N.}~\bibnamefont {Srivastava}}, \bibinfo {author} {\bibfnamefont {G.}~\bibnamefont {Hinton}}, \bibinfo {author} {\bibfnamefont {A.}~\bibnamefont {Krizhevsky}}, \bibinfo {author} {\bibfnamefont {I.}~\bibnamefont {Sutskever}},\ and\ \bibinfo {author} {\bibfnamefont {R.}~\bibnamefont {Salakhutdinov}},\ }\bibfield  {title} {\bibinfo {title} {Dropout: {{A Simple Way}} to {{Prevent Neural Networks}} from {{Overfitting}}},\ }\href {http://jmlr.org/papers/v15/srivastava14a.html} {\bibfield  {journal} {\bibinfo  {journal} {Journal of Machine Learning Research}\ }\textbf {\bibinfo {volume} {15}},\ \bibinfo {pages} {1929} (\bibinfo {year} {2014})}\BibitemShut {NoStop}%
\bibitem [{\citenamefont {Wan}\ \emph {et~al.}(2013)\citenamefont {Wan}, \citenamefont {Zeiler}, \citenamefont {Zhang}, \citenamefont {Cun},\ and\ \citenamefont {Fergus}}]{wan2013regularization}%
  \BibitemOpen
  \bibfield  {author} {\bibinfo {author} {\bibfnamefont {L.}~\bibnamefont {Wan}}, \bibinfo {author} {\bibfnamefont {M.}~\bibnamefont {Zeiler}}, \bibinfo {author} {\bibfnamefont {S.}~\bibnamefont {Zhang}}, \bibinfo {author} {\bibfnamefont {Y.~L.}\ \bibnamefont {Cun}},\ and\ \bibinfo {author} {\bibfnamefont {R.}~\bibnamefont {Fergus}},\ }\bibfield  {title} {\bibinfo {title} {Regularization of {{Neural Networks}} using {{DropConnect}}},\ }in\ \href {https://proceedings.mlr.press/v28/wan13.html} {\emph {\bibinfo {booktitle} {Proceedings of the 30th {{International Conference}} on {{Machine Learning}}}}}\ (\bibinfo  {publisher} {PMLR},\ \bibinfo {year} {2013})\ pp.\ \bibinfo {pages} {1058--1066}\BibitemShut {NoStop}%
\bibitem [{\citenamefont {Frankle}\ and\ \citenamefont {Carbin}(2018)}]{frankle2018lottery}%
  \BibitemOpen
  \bibfield  {author} {\bibinfo {author} {\bibfnamefont {J.}~\bibnamefont {Frankle}}\ and\ \bibinfo {author} {\bibfnamefont {M.}~\bibnamefont {Carbin}},\ }\bibfield  {title} {\bibinfo {title} {The {{Lottery Ticket Hypothesis}}: {{Finding Sparse}}, {{Trainable Neural Networks}}},\ }in\ \href {https://openreview.net/forum?id=rJl-b3RcF7} {\emph {\bibinfo {booktitle} {International {{Conference}} on {{Learning Representations}}}}}\ (\bibinfo {year} {2018})\BibitemShut {NoStop}%
\bibitem [{\citenamefont {Malach}\ \emph {et~al.}(2020)\citenamefont {Malach}, \citenamefont {Yehudai}, \citenamefont {{Shalev-shwartz}},\ and\ \citenamefont {Shamir}}]{malach2020proving}%
  \BibitemOpen
  \bibfield  {author} {\bibinfo {author} {\bibfnamefont {E.}~\bibnamefont {Malach}}, \bibinfo {author} {\bibfnamefont {G.}~\bibnamefont {Yehudai}}, \bibinfo {author} {\bibfnamefont {S.}~\bibnamefont {{Shalev-shwartz}}},\ and\ \bibinfo {author} {\bibfnamefont {O.}~\bibnamefont {Shamir}},\ }\bibfield  {title} {\bibinfo {title} {Proving the lottery ticket hypothesis: Pruning is all you need},\ }in\ \href {https://dl.acm.org/doi/10.5555/3524938.3525558} {\emph {\bibinfo {booktitle} {Proceedings of the 37th {{International Conference}} on {{Machine Learning}}}}},\ \bibinfo {series} {{{ICML}}'20}, Vol.\ \bibinfo {volume} {119}\ (\bibinfo  {publisher} {JMLR.org},\ \bibinfo {year} {2020})\ pp.\ \bibinfo {pages} {6682--6691}\BibitemShut {NoStop}%
\bibitem [{\citenamefont {Du}\ \emph {et~al.}(2019)\citenamefont {Du}, \citenamefont {Lee}, \citenamefont {Li}, \citenamefont {Wang},\ and\ \citenamefont {Zhai}}]{du2019gradient}%
  \BibitemOpen
  \bibfield  {author} {\bibinfo {author} {\bibfnamefont {S.}~\bibnamefont {Du}}, \bibinfo {author} {\bibfnamefont {J.}~\bibnamefont {Lee}}, \bibinfo {author} {\bibfnamefont {H.}~\bibnamefont {Li}}, \bibinfo {author} {\bibfnamefont {L.}~\bibnamefont {Wang}},\ and\ \bibinfo {author} {\bibfnamefont {X.}~\bibnamefont {Zhai}},\ }\bibfield  {title} {\bibinfo {title} {Gradient {{Descent Finds Global Minima}} of {{Deep Neural Networks}}},\ }in\ \href {https://proceedings.mlr.press/v97/du19c.html} {\emph {\bibinfo {booktitle} {Proceedings of the 36th {{International Conference}} on {{Machine Learning}}}}}\ (\bibinfo  {publisher} {PMLR},\ \bibinfo {year} {2019})\ pp.\ \bibinfo {pages} {1675--1685}\BibitemShut {NoStop}%
\bibitem [{\citenamefont {{Allen-Zhu}}\ \emph {et~al.}(2019)\citenamefont {{Allen-Zhu}}, \citenamefont {Li},\ and\ \citenamefont {Song}}]{allen-zhu2019convergence}%
  \BibitemOpen
  \bibfield  {author} {\bibinfo {author} {\bibfnamefont {Z.}~\bibnamefont {{Allen-Zhu}}}, \bibinfo {author} {\bibfnamefont {Y.}~\bibnamefont {Li}},\ and\ \bibinfo {author} {\bibfnamefont {Z.}~\bibnamefont {Song}},\ }\bibfield  {title} {\bibinfo {title} {A {{Convergence Theory}} for {{Deep Learning}} via {{Over-Parameterization}}},\ }in\ \href {https://proceedings.mlr.press/v97/allen-zhu19a.html} {\emph {\bibinfo {booktitle} {Proceedings of the 36th {{International Conference}} on {{Machine Learning}}}}}\ (\bibinfo  {publisher} {PMLR},\ \bibinfo {year} {2019})\ pp.\ \bibinfo {pages} {242--252}\BibitemShut {NoStop}%
\bibitem [{\citenamefont {Zhu}\ and\ \citenamefont {Gupta}(2018)}]{zhu2018prune}%
  \BibitemOpen
  \bibfield  {author} {\bibinfo {author} {\bibfnamefont {M.~H.}\ \bibnamefont {Zhu}}\ and\ \bibinfo {author} {\bibfnamefont {S.}~\bibnamefont {Gupta}},\ }\bibfield  {title} {\bibinfo {title} {To {{Prune}}, or {{Not}} to {{Prune}}: {{Exploring}} the {{Efficacy}} of {{Pruning}} for {{Model Compression}}},\ }\href {https://openreview.net/forum?id=Sy1iIDkPM} {\  (\bibinfo {year} {2018})}\BibitemShut {NoStop}%
\bibitem [{\citenamefont {Liu}\ \emph {et~al.}(2018)\citenamefont {Liu}, \citenamefont {Sun}, \citenamefont {Zhou}, \citenamefont {Huang},\ and\ \citenamefont {Darrell}}]{liu2018rethinking}%
  \BibitemOpen
  \bibfield  {author} {\bibinfo {author} {\bibfnamefont {Z.}~\bibnamefont {Liu}}, \bibinfo {author} {\bibfnamefont {M.}~\bibnamefont {Sun}}, \bibinfo {author} {\bibfnamefont {T.}~\bibnamefont {Zhou}}, \bibinfo {author} {\bibfnamefont {G.}~\bibnamefont {Huang}},\ and\ \bibinfo {author} {\bibfnamefont {T.}~\bibnamefont {Darrell}},\ }\bibfield  {title} {\bibinfo {title} {Rethinking the {{Value}} of {{Network Pruning}}},\ }in\ \href {https://openreview.net/forum?id=rJlnB3C5Ym} {\emph {\bibinfo {booktitle} {International {{Conference}} on {{Learning Representations}}}}}\ (\bibinfo {year} {2018})\BibitemShut {NoStop}%
\bibitem [{\citenamefont {Zhang}\ \emph {et~al.}(2024)\citenamefont {Zhang}, \citenamefont {Zhang}, \citenamefont {Sun},\ and\ \citenamefont {Liu}}]{zhang2024how}%
  \BibitemOpen
  \bibfield  {author} {\bibinfo {author} {\bibfnamefont {Q.}~\bibnamefont {Zhang}}, \bibinfo {author} {\bibfnamefont {R.}~\bibnamefont {Zhang}}, \bibinfo {author} {\bibfnamefont {J.}~\bibnamefont {Sun}},\ and\ \bibinfo {author} {\bibfnamefont {Y.}~\bibnamefont {Liu}},\ }\bibfield  {title} {\bibinfo {title} {How {{Sparse Can We Prune A Deep Network}}: {{A Fundamental Limit Perspective}}},\ }in\ \href {https://openreview.net/forum?id=IAAPhOLhcX&referrer=%5Bthe%20profile%20of%20Qiaozhe%20Zhang%5D(%2Fprofile%3Fid%3D~Qiaozhe_Zhang1)} {\emph {\bibinfo {booktitle} {The {{Thirty-eighth Annual Conference}} on {{Neural Information Processing Systems}}}}}\ (\bibinfo {year} {2024})\BibitemShut {NoStop}%
\bibitem [{\citenamefont {Evci}\ \emph {et~al.}(2020)\citenamefont {Evci}, \citenamefont {Gale}, \citenamefont {Menick}, \citenamefont {Castro},\ and\ \citenamefont {Elsen}}]{evci2020rigging}%
  \BibitemOpen
  \bibfield  {author} {\bibinfo {author} {\bibfnamefont {U.}~\bibnamefont {Evci}}, \bibinfo {author} {\bibfnamefont {T.}~\bibnamefont {Gale}}, \bibinfo {author} {\bibfnamefont {J.}~\bibnamefont {Menick}}, \bibinfo {author} {\bibfnamefont {P.~S.}\ \bibnamefont {Castro}},\ and\ \bibinfo {author} {\bibfnamefont {E.}~\bibnamefont {Elsen}},\ }\bibfield  {title} {\bibinfo {title} {Rigging the lottery: Making all tickets winners},\ }in\ \href {https://dl.acm.org/doi/10.5555/3524938.3525214} {\emph {\bibinfo {booktitle} {Proceedings of the 37th {{International Conference}} on {{Machine Learning}}}}},\ \bibinfo {series} {{{ICML}}'20}, Vol.\ \bibinfo {volume} {119}\ (\bibinfo  {publisher} {JMLR.org},\ \bibinfo {year} {2020})\ pp.\ \bibinfo {pages} {2943--2952}\BibitemShut {NoStop}%
\bibitem [{\citenamefont {Seung}\ \emph {et~al.}(1992)\citenamefont {Seung}, \citenamefont {Sompolinsky},\ and\ \citenamefont {Tishby}}]{seung1992statistical}%
  \BibitemOpen
  \bibfield  {author} {\bibinfo {author} {\bibfnamefont {H.~S.}\ \bibnamefont {Seung}}, \bibinfo {author} {\bibfnamefont {H.}~\bibnamefont {Sompolinsky}},\ and\ \bibinfo {author} {\bibfnamefont {N.}~\bibnamefont {Tishby}},\ }\bibfield  {title} {\bibinfo {title} {Statistical mechanics of learning from examples},\ }\href {https://link.aps.org/doi/10.1103/PhysRevA.45.6056} {\bibfield  {journal} {\bibinfo  {journal} {Phys. Rev. A}\ }\textbf {\bibinfo {volume} {45}},\ \bibinfo {pages} {6056} (\bibinfo {year} {1992})}\BibitemShut {NoStop}%
\bibitem [{\citenamefont {Hestness}\ \emph {et~al.}(2017)\citenamefont {Hestness}, \citenamefont {Narang}, \citenamefont {Ardalani}, \citenamefont {Diamos}, \citenamefont {Jun}, \citenamefont {Kianinejad}, \citenamefont {Patwary}, \citenamefont {Yang},\ and\ \citenamefont {Zhou}}]{hestness2017deep}%
  \BibitemOpen
  \bibfield  {author} {\bibinfo {author} {\bibfnamefont {J.}~\bibnamefont {Hestness}}, \bibinfo {author} {\bibfnamefont {S.}~\bibnamefont {Narang}}, \bibinfo {author} {\bibfnamefont {N.}~\bibnamefont {Ardalani}}, \bibinfo {author} {\bibfnamefont {G.}~\bibnamefont {Diamos}}, \bibinfo {author} {\bibfnamefont {H.}~\bibnamefont {Jun}}, \bibinfo {author} {\bibfnamefont {H.}~\bibnamefont {Kianinejad}}, \bibinfo {author} {\bibfnamefont {M.~M.~A.}\ \bibnamefont {Patwary}}, \bibinfo {author} {\bibfnamefont {Y.}~\bibnamefont {Yang}},\ and\ \bibinfo {author} {\bibfnamefont {Y.}~\bibnamefont {Zhou}},\ }\href {http://arxiv.org/abs/1712.00409} {\bibinfo {title} {Deep {{Learning Scaling}} is {{Predictable}}, {{Empirically}}}} (\bibinfo {year} {2017})\BibitemShut {NoStop}%
\bibitem [{\citenamefont {Spigler}\ \emph {et~al.}(2020)\citenamefont {Spigler}, \citenamefont {Geiger},\ and\ \citenamefont {Wyart}}]{spigler2020asymptotic}%
  \BibitemOpen
  \bibfield  {author} {\bibinfo {author} {\bibfnamefont {S.}~\bibnamefont {Spigler}}, \bibinfo {author} {\bibfnamefont {M.}~\bibnamefont {Geiger}},\ and\ \bibinfo {author} {\bibfnamefont {M.}~\bibnamefont {Wyart}},\ }\bibfield  {title} {\bibinfo {title} {Asymptotic learning curves of kernel methods: Empirical data versus teacher--student paradigm},\ }\href {https://doi.org/10.1088/1742-5468/abc61d} {\bibfield  {journal} {\bibinfo  {journal} {J. Stat. Mech.}\ }\textbf {\bibinfo {volume} {2020}},\ \bibinfo {pages} {124001} (\bibinfo {year} {2020})}\BibitemShut {NoStop}%
\bibitem [{\citenamefont {Kaplan}\ \emph {et~al.}(2020)\citenamefont {Kaplan}, \citenamefont {McCandlish}, \citenamefont {Henighan}, \citenamefont {Brown}, \citenamefont {Chess}, \citenamefont {Child}, \citenamefont {Gray}, \citenamefont {Radford}, \citenamefont {Wu},\ and\ \citenamefont {Amodei}}]{kaplan2020scaling}%
  \BibitemOpen
  \bibfield  {author} {\bibinfo {author} {\bibfnamefont {J.}~\bibnamefont {Kaplan}}, \bibinfo {author} {\bibfnamefont {S.}~\bibnamefont {McCandlish}}, \bibinfo {author} {\bibfnamefont {T.}~\bibnamefont {Henighan}}, \bibinfo {author} {\bibfnamefont {T.~B.}\ \bibnamefont {Brown}}, \bibinfo {author} {\bibfnamefont {B.}~\bibnamefont {Chess}}, \bibinfo {author} {\bibfnamefont {R.}~\bibnamefont {Child}}, \bibinfo {author} {\bibfnamefont {S.}~\bibnamefont {Gray}}, \bibinfo {author} {\bibfnamefont {A.}~\bibnamefont {Radford}}, \bibinfo {author} {\bibfnamefont {J.}~\bibnamefont {Wu}},\ and\ \bibinfo {author} {\bibfnamefont {D.}~\bibnamefont {Amodei}},\ }\href {http://arxiv.org/abs/2001.08361} {\bibinfo {title} {Scaling {{Laws}} for {{Neural Language Models}}}} (\bibinfo {year} {2020})\BibitemShut {NoStop}%
\bibitem [{\citenamefont {Rosenfeld}\ \emph {et~al.}(2020)\citenamefont {Rosenfeld}, \citenamefont {Rosenfeld}, \citenamefont {Belinkov},\ and\ \citenamefont {Shavit}}]{rosenfeld2020constructive}%
  \BibitemOpen
  \bibfield  {author} {\bibinfo {author} {\bibfnamefont {J.~S.}\ \bibnamefont {Rosenfeld}}, \bibinfo {author} {\bibfnamefont {A.}~\bibnamefont {Rosenfeld}}, \bibinfo {author} {\bibfnamefont {Y.}~\bibnamefont {Belinkov}},\ and\ \bibinfo {author} {\bibfnamefont {N.}~\bibnamefont {Shavit}},\ }\bibfield  {title} {\bibinfo {title} {A {{Constructive Prediction}} of the {{Generalization Error Across Scales}}},\ }in\ \href {https://iclr.cc/virtual_2020/poster_ryenvpEKDr.html} {\emph {\bibinfo {booktitle} {Eighth {{International Conference}} on {{Learning Representations}}}}}\ (\bibinfo {year} {2020})\BibitemShut {NoStop}%
\bibitem [{\citenamefont {Bahri}\ \emph {et~al.}(2020)\citenamefont {Bahri}, \citenamefont {Kadmon}, \citenamefont {Pennington}, \citenamefont {Schoenholz}, \citenamefont {{Sohl-Dickstein}},\ and\ \citenamefont {Ganguli}}]{bahri2020statistical}%
  \BibitemOpen
  \bibfield  {author} {\bibinfo {author} {\bibfnamefont {Y.}~\bibnamefont {Bahri}}, \bibinfo {author} {\bibfnamefont {J.}~\bibnamefont {Kadmon}}, \bibinfo {author} {\bibfnamefont {J.}~\bibnamefont {Pennington}}, \bibinfo {author} {\bibfnamefont {S.~S.}\ \bibnamefont {Schoenholz}}, \bibinfo {author} {\bibfnamefont {J.}~\bibnamefont {{Sohl-Dickstein}}},\ and\ \bibinfo {author} {\bibfnamefont {S.}~\bibnamefont {Ganguli}},\ }\bibfield  {title} {\bibinfo {title} {Statistical {{Mechanics}} of {{Deep Learning}}},\ }\href {https://www.annualreviews.org/content/journals/10.1146/annurev-conmatphys-031119-050745} {\bibfield  {journal} {\bibinfo  {journal} {Annual Review of Condensed Matter Physics}\ }\textbf {\bibinfo {volume} {11}},\ \bibinfo {pages} {501} (\bibinfo {year} {2020})}\BibitemShut {NoStop}%
\bibitem [{\citenamefont {Bordelon}\ \emph {et~al.}(2020)\citenamefont {Bordelon}, \citenamefont {Canatar},\ and\ \citenamefont {Pehlevan}}]{bordelon2020spectrum}%
  \BibitemOpen
  \bibfield  {author} {\bibinfo {author} {\bibfnamefont {B.}~\bibnamefont {Bordelon}}, \bibinfo {author} {\bibfnamefont {A.}~\bibnamefont {Canatar}},\ and\ \bibinfo {author} {\bibfnamefont {C.}~\bibnamefont {Pehlevan}},\ }\bibfield  {title} {\bibinfo {title} {Spectrum {{Dependent Learning Curves}} in {{Kernel Regression}} and {{Wide Neural Networks}}},\ }in\ \href {https://proceedings.mlr.press/v119/bordelon20a.html} {\emph {\bibinfo {booktitle} {Proceedings of the 37th {{International Conference}} on {{Machine Learning}}}}}\ (\bibinfo  {publisher} {PMLR},\ \bibinfo {year} {2020})\ pp.\ \bibinfo {pages} {1024--1034}\BibitemShut {NoStop}%
\bibitem [{\citenamefont {Bengio}\ \emph {et~al.}(1994)\citenamefont {Bengio}, \citenamefont {Simard},\ and\ \citenamefont {Frasconi}}]{bengio1994learning}%
  \BibitemOpen
  \bibfield  {author} {\bibinfo {author} {\bibfnamefont {Y.}~\bibnamefont {Bengio}}, \bibinfo {author} {\bibfnamefont {P.}~\bibnamefont {Simard}},\ and\ \bibinfo {author} {\bibfnamefont {P.}~\bibnamefont {Frasconi}},\ }\bibfield  {title} {\bibinfo {title} {Learning long-term dependencies with gradient descent is difficult},\ }\href {https://ieeexplore.ieee.org/abstract/document/279181} {\bibfield  {journal} {\bibinfo  {journal} {IEEE Transactions on Neural Networks}\ }\textbf {\bibinfo {volume} {5}},\ \bibinfo {pages} {157} (\bibinfo {year} {1994})}\BibitemShut {NoStop}%
\bibitem [{\citenamefont {Hochreiter}(1998)}]{hochreiter1998vanishing}%
  \BibitemOpen
  \bibfield  {author} {\bibinfo {author} {\bibfnamefont {S.}~\bibnamefont {Hochreiter}},\ }\bibfield  {title} {\bibinfo {title} {The {{Vanishing Gradient Problem During Learning Recurrent Neural Nets}} and {{Problem Solutions}}},\ }\href {https://www.worldscientific.com/doi/abs/10.1142/s0218488598000094} {\bibfield  {journal} {\bibinfo  {journal} {Int. J. Unc. Fuzz. Knowl. Based Syst.}\ }\textbf {\bibinfo {volume} {06}},\ \bibinfo {pages} {107} (\bibinfo {year} {1998})}\BibitemShut {NoStop}%
\bibitem [{\citenamefont {Glorot}\ and\ \citenamefont {Bengio}(2010)}]{glorot2010understanding}%
  \BibitemOpen
  \bibfield  {author} {\bibinfo {author} {\bibfnamefont {X.}~\bibnamefont {Glorot}}\ and\ \bibinfo {author} {\bibfnamefont {Y.}~\bibnamefont {Bengio}},\ }\bibfield  {title} {\bibinfo {title} {Understanding the difficulty of training deep feedforward neural networks},\ }in\ \href {https://proceedings.mlr.press/v9/glorot10a.html} {\emph {\bibinfo {booktitle} {Proceedings of the {{Thirteenth International Conference}} on {{Artificial Intelligence}} and {{Statistics}}}}}\ (\bibinfo  {publisher} {{JMLR Workshop and Conference Proceedings}},\ \bibinfo {year} {2010})\ pp.\ \bibinfo {pages} {249--256}\BibitemShut {NoStop}%
\bibitem [{\citenamefont {He}\ \emph {et~al.}(2016)\citenamefont {He}, \citenamefont {Zhang}, \citenamefont {Ren},\ and\ \citenamefont {Sun}}]{he2016deep}%
  \BibitemOpen
  \bibfield  {author} {\bibinfo {author} {\bibfnamefont {K.}~\bibnamefont {He}}, \bibinfo {author} {\bibfnamefont {X.}~\bibnamefont {Zhang}}, \bibinfo {author} {\bibfnamefont {S.}~\bibnamefont {Ren}},\ and\ \bibinfo {author} {\bibfnamefont {J.}~\bibnamefont {Sun}},\ }\bibfield  {title} {\bibinfo {title} {Deep {{Residual Learning}} for {{Image Recognition}}},\ }in\ \href {https://openaccess.thecvf.com/content_cvpr_2016/html/He_Deep_Residual_Learning_CVPR_2016_paper.html} {\emph {\bibinfo {booktitle} {Proceedings of the {{IEEE Conference}} on {{Computer Vision}} and {{Pattern Recognition}}}}}\ (\bibinfo {year} {2016})\ pp.\ \bibinfo {pages} {770--778}\BibitemShut {NoStop}%
\bibitem [{\citenamefont {Lecun}\ \emph {et~al.}(1998)\citenamefont {Lecun}, \citenamefont {Bottou}, \citenamefont {Bengio},\ and\ \citenamefont {Haffner}}]{lecun1998gradientbased}%
  \BibitemOpen
  \bibfield  {author} {\bibinfo {author} {\bibfnamefont {Y.}~\bibnamefont {Lecun}}, \bibinfo {author} {\bibfnamefont {L.}~\bibnamefont {Bottou}}, \bibinfo {author} {\bibfnamefont {Y.}~\bibnamefont {Bengio}},\ and\ \bibinfo {author} {\bibfnamefont {P.}~\bibnamefont {Haffner}},\ }\bibfield  {title} {\bibinfo {title} {Gradient-based learning applied to document recognition},\ }\href {https://ieeexplore.ieee.org/document/726791} {\bibfield  {journal} {\bibinfo  {journal} {Proceedings of the IEEE}\ }\textbf {\bibinfo {volume} {86}},\ \bibinfo {pages} {2278} (\bibinfo {year} {1998})}\BibitemShut {NoStop}%
\bibitem [{\citenamefont {Kingma}\ and\ \citenamefont {Ba}(2015)}]{kingma2015adam}%
  \BibitemOpen
  \bibfield  {author} {\bibinfo {author} {\bibfnamefont {D.~P.}\ \bibnamefont {Kingma}}\ and\ \bibinfo {author} {\bibfnamefont {J.}~\bibnamefont {Ba}},\ }\bibfield  {title} {\bibinfo {title} {Adam: A method for stochastic optimization},\ }in\ \href {https://arxiv.org/abs/1412.6980} {\emph {\bibinfo {booktitle} {International Conference on Learning Representations ({{ICLR}})}}}\ (\bibinfo {year} {2015})\BibitemShut {NoStop}%
\bibitem [{\citenamefont {Gal}\ and\ \citenamefont {Ghahramani}(2016)}]{gal2016dropout}%
  \BibitemOpen
  \bibfield  {author} {\bibinfo {author} {\bibfnamefont {Y.}~\bibnamefont {Gal}}\ and\ \bibinfo {author} {\bibfnamefont {Z.}~\bibnamefont {Ghahramani}},\ }\bibfield  {title} {\bibinfo {title} {Dropout as a {{Bayesian Approximation}}: {{Representing Model Uncertainty}} in {{Deep Learning}}},\ }in\ \href {https://proceedings.mlr.press/v48/gal16.html} {\emph {\bibinfo {booktitle} {Proceedings of {{The}} 33rd {{International Conference}} on {{Machine Learning}}}}}\ (\bibinfo  {publisher} {PMLR},\ \bibinfo {year} {2016})\ pp.\ \bibinfo {pages} {1050--1059}\BibitemShut {NoStop}%
\bibitem [{\citenamefont {Berezinskii}(1971)}]{berezinskii1971destruction}%
  \BibitemOpen
  \bibfield  {author} {\bibinfo {author} {\bibfnamefont {V.~L.}\ \bibnamefont {Berezinskii}},\ }\bibfield  {title} {\bibinfo {title} {Destruction of {{Long-range Order}} in {{One-dimensional}} and {{Two-dimensional Systems}} having a {{Continuous Symmetry Group I}}. {{Classical Systems}}},\ }\href@noop {} {\bibfield  {journal} {\bibinfo  {journal} {Sov. Phys. JETP}\ }\textbf {\bibinfo {volume} {32}},\ \bibinfo {pages} {493} (\bibinfo {year} {1971})}\BibitemShut {NoStop}%
\bibitem [{\citenamefont {Kosterlitz}\ and\ \citenamefont {Thouless}(1973)}]{kosterlitz1973ordering}%
  \BibitemOpen
  \bibfield  {author} {\bibinfo {author} {\bibfnamefont {J.~M.}\ \bibnamefont {Kosterlitz}}\ and\ \bibinfo {author} {\bibfnamefont {D.~J.}\ \bibnamefont {Thouless}},\ }\bibfield  {title} {\bibinfo {title} {Ordering, metastability and phase transitions in two-dimensional systems},\ }\href {https://iopscience.iop.org/article/10.1088/0022-3719/6/7/010} {\bibfield  {journal} {\bibinfo  {journal} {J. Phys. C: Solid State Phys.}\ }\textbf {\bibinfo {volume} {6}},\ \bibinfo {pages} {1181} (\bibinfo {year} {1973})}\BibitemShut {NoStop}%
\bibitem [{\citenamefont {Kosterlitz}(1974)}]{kosterlitz1974critical}%
  \BibitemOpen
  \bibfield  {author} {\bibinfo {author} {\bibfnamefont {J.~M.}\ \bibnamefont {Kosterlitz}},\ }\bibfield  {title} {\bibinfo {title} {The critical properties of the two-dimensional xy model},\ }\href {https://iopscience.iop.org/article/10.1088/0022-3719/7/6/005} {\bibfield  {journal} {\bibinfo  {journal} {J. Phys. C: Solid State Phys.}\ }\textbf {\bibinfo {volume} {7}},\ \bibinfo {pages} {1046} (\bibinfo {year} {1974})}\BibitemShut {NoStop}%
\bibitem [{\citenamefont {Kosterlitz}(2016)}]{kosterlitz2016kosterlitz}%
  \BibitemOpen
  \bibfield  {author} {\bibinfo {author} {\bibfnamefont {J.~M.}\ \bibnamefont {Kosterlitz}},\ }\bibfield  {title} {\bibinfo {title} {Kosterlitz--{{Thouless}} physics: A review of key issues},\ }\href {https://iopscience.iop.org/article/10.1088/0034-4885/79/2/026001} {\bibfield  {journal} {\bibinfo  {journal} {Rep. Prog. Phys.}\ }\textbf {\bibinfo {volume} {79}},\ \bibinfo {pages} {026001} (\bibinfo {year} {2016})}\BibitemShut {NoStop}%
\bibitem [{\citenamefont {Pan}(2025)}]{pan2025fss}%
  \BibitemOpen
  \bibfield  {author} {\bibinfo {author} {\bibfnamefont {H.}~\bibnamefont {Pan}},\ }\href {https://github.com/Pixley-Research-Group-in-CMT/FSS} {\bibinfo {title} {{{FSS}}: {{Finite-size}} scaling toolkit}} (\bibinfo {year} {2025})\BibitemShut {NoStop}%
\bibitem [{\citenamefont {Huttenlocher}(1979)}]{huttenlocher1979synaptic}%
  \BibitemOpen
  \bibfield  {author} {\bibinfo {author} {\bibfnamefont {P.~R.}\ \bibnamefont {Huttenlocher}},\ }\bibfield  {title} {\bibinfo {title} {Synaptic density in human frontal cortex - developmental changes and effects of aging},\ }\href {https://doi.org/10.1016/0006-8993(79)90349-4} {\bibfield  {journal} {\bibinfo  {journal} {Brain Res}\ }\textbf {\bibinfo {volume} {163}},\ \bibinfo {pages} {195} (\bibinfo {year} {1979})}\BibitemShut {NoStop}%
\bibitem [{\citenamefont {Chechik}\ \emph {et~al.}(1998)\citenamefont {Chechik}, \citenamefont {Meilijson},\ and\ \citenamefont {Ruppin}}]{chechik1998synaptic}%
  \BibitemOpen
  \bibfield  {author} {\bibinfo {author} {\bibfnamefont {G.}~\bibnamefont {Chechik}}, \bibinfo {author} {\bibfnamefont {I.}~\bibnamefont {Meilijson}},\ and\ \bibinfo {author} {\bibfnamefont {E.}~\bibnamefont {Ruppin}},\ }\bibfield  {title} {\bibinfo {title} {Synaptic pruning in development: A computational account},\ }\href {https://doi.org/10.1162/089976698300017124} {\bibfield  {journal} {\bibinfo  {journal} {Neural Comput}\ }\textbf {\bibinfo {volume} {10}},\ \bibinfo {pages} {1759} (\bibinfo {year} {1998})}\BibitemShut {NoStop}%
\bibitem [{\citenamefont {Navlakha}\ \emph {et~al.}(2015)\citenamefont {Navlakha}, \citenamefont {Barth},\ and\ \citenamefont {{Bar-Joseph}}}]{navlakha2015decreasingrate}%
  \BibitemOpen
  \bibfield  {author} {\bibinfo {author} {\bibfnamefont {S.}~\bibnamefont {Navlakha}}, \bibinfo {author} {\bibfnamefont {A.~L.}\ \bibnamefont {Barth}},\ and\ \bibinfo {author} {\bibfnamefont {Z.}~\bibnamefont {{Bar-Joseph}}},\ }\bibfield  {title} {\bibinfo {title} {Decreasing-{{Rate Pruning Optimizes}} the {{Construction}} of {{Efficient}} and {{Robust Distributed Networks}}},\ }\href {https://doi.org/10.1371/journal.pcbi.1004347} {\bibfield  {journal} {\bibinfo  {journal} {PLoS Comput Biol}\ }\textbf {\bibinfo {volume} {11}},\ \bibinfo {pages} {e1004347} (\bibinfo {year} {2015})}\BibitemShut {NoStop}%
\end{thebibliography}%

\clearpage
\appendix
% \begin{table}[h]
% \centering
% \begin{tabular}{l l}
% \hline
% Symbol & Description \\
% \hline
% $p_{\text{train}}$ & train dropout rate \\
% $p_{\text{eval}}$ & evaluation dropout rate \\
% $N$ & training sample size \\
% $n_h$ & width \\
% $L$ & number of hidden layers \\
% $\mathcal{L}_{\text{CE}}$ & cross entropy loss \\
% $\mathcal{A}$ & accuracy \\
% $\alpha_{\text{CE}}$ & slope of the cross entropy\\
% $\alpha_{\mathcal{A}}$ & slope of the accuracy\\
% $\phi$ & activation function \\
% $N_{\text{test}}$ & number of test samples \\
% $\expval{\cdot}$ & average over stochasticity \\
% \hline
% \end{tabular}
% \caption{Table of notations: Internal use only}
% \end{table}
\section{finite-size scaling using BKT-like transition ansatz}\label{app:bkt}
\para{Derive the finite-size scaling ansatz for the BKT-like transition, Eq.~\eqref{eq:bkt} }
In the two-dimensional XY model, the BKT transition~\cite{berezinskii1971destruction, kosterlitz1973ordering, kosterlitz1974critical} separates a low-temperature phase with QLRO from a high-temperature disordered phase. On the disordered side, the correlation length diverges as
\begin{equation}
\xi_{\text{BKT}} \sim \exp\left(\frac{a}{\sqrt{T - T_{\text{BKT}}}}\right), \quad T \to T_{\text{BKT}}^+,
\end{equation}
where $a$ is a constant. The square-root exponent follows from the renormalization group flow of the vortex fugacity~\cite{kosterlitz1974critical}. This essential singularity distinguishes the BKT transition from conventional critical points, where the correlation length diverges algebraically as $\xi \sim \abs{T_c - T}^{-\nu}$.

As argued in Sec.~\ref{sec:bkt}, the power-law scaling of the cross-entropy loss with training dataset size $N$ in the eumentia phase [Eq.~\eqref{eq:CE}] is the analog of QLRO, with the evaluation dropout rate $p_{\text{eval}}$ playing the role of temperature. Since our identification is phenomenological rather than based on a microscopic derivation, we generalize the stretched exponential form to
\begin{equation}
\xi \sim \exp\left(\frac{a}{\abs{p - p_c}^{\sigma}}\right),
\end{equation}
where the exponent $\sigma$ is a free parameter that we extract from the data, recovering the standard BKT value when $\sigma = 1/2$. We now derive the finite-size scaling ansatz associated with this correlation length.

In the standard theory of critical phenomena, the finite-size scaling hypothesis states that near a continuous phase transition, singular quantities in a system of finite size $L$ are universal functions of $L/\xi$, where $\xi$ is the correlation length. The finite-size effects become important when $\xi$ becomes comparable to $L$. Setting $\xi \sim L$ and taking the logarithm,
\begin{equation}
\log(L/L_0) \sim \frac{a}{\abs{p - p_c}^{\sigma}},
\end{equation}
where $L_0$ is a microscopic reference scale. At the crossing,
\begin{equation}
\abs{p - p_c} \sim \left[\log(L/L_0)\right]^{-1/\sigma},
\end{equation}
so the critical region shrinks logarithmically with system size, much slower than the algebraic shrinkage $\sim L^{-1/\nu}$ in a conventional transition (see Appendix~\ref{app:powerlaw}). The natural scaling variable is the distance from criticality measured in units of this crossover width,
\begin{equation}
x = (p - p_c) \left[\log(L/L_0)\right]^{1/\sigma},
\end{equation}
and the finite-size scaling ansatz for any singular quantity takes the form $f(x)$, where $f$ is a universal scaling function.

In the neural network, the training dataset size $N$ plays the role of the system size $L$, and the evaluation dropout rate $p_{\text{eval}}$ serves as the tuning parameter. The cross-entropy loss $\expval{\mathcal{L}_{\text{CE}}}$, already an intensive quantity averaged over the test set, does not carry an anomalous scaling prefactor. The BKT finite-size scaling ansatz thus becomes
\begin{equation}
\expval{\mathcal{L}_{\text{CE}}} \sim f\left((p_{\text{eval}} - p_{\text{eval}, c}) \left[\log(N/N_0)\right]^{1/\sigma}\right),
\end{equation}
which is Eq.~\eqref{eq:bkt} in the main text, where $p_{\text{eval}, c}$ is the critical evaluation dropout rate and $N_0$ is a microscopic scale fixed to $N_0 = 1$ as discussed in Sec.~\ref{sec:bkt}.

\para{Define $\chi_\nu^2$ for the goodness of the data collapse}
To quantify the quality of the data collapse, we define the reduced chi-squared $\chi_\nu^2$ as follows. For a given choice of fitting parameters $(\sigma, p_{\text{eval},c})$, we compute the scaling variable
\begin{equation}
x_i = (p_{\text{eval},i} - p_{\text{eval},c}) \left[\log(N_i/N_0)\right]^{1/\sigma}
\end{equation}
for each data point $i$ and sort all data points in ascending order of $x_i$. Denoting the corresponding cross-entropy loss as $y_i \equiv \expval{\mathcal{L}_{\text{CE}}}_i$, we compare each point to the linear interpolation between its two sorted neighbors,
\begin{equation}
y_i' = y_{i-1} + \frac{y_{i+1} - y_{i-1}}{x_{i+1} - x_{i-1}}(x_i - x_{i-1}),
\end{equation}
with one-sided interpolation at the two endpoints. The reduced chi-squared is then
\begin{equation}\label{eq:chi2}
\chi_\nu^2 = \frac{1}{\mathcal{N} - n_f} \sum_{i=1}^{\mathcal{N}} \left(\frac{y_i - y_i'}{s_i}\right)^2,
\end{equation}
where $\mathcal{N}$ is the total number of data points, $n_f$ is the number of fitting parameters ($n_f = 2$ for $\sigma$ and $p_{\text{eval},c}$), $\nu=\mathcal{N}-n_f$ is the total degrees of freedom and $s_i$ is the combined uncertainty accounting for error propagation through the interpolation,
\begin{equation}
s_i^2 = \delta_{y_i}^2 + \left(\frac{x_{i+1} - x_i}{x_{i+1} - x_{i-1}}\right)^2 \delta_{y_{i-1}}^2 + \left(\frac{x_i - x_{i-1}}{x_{i+1} - x_{i-1}}\right)^2 \delta_{y_{i+1}}^2.
\end{equation}
Here, $\delta_{y_i}$ is the standard error of the mean of $y_i$, estimated using the bootstrap method: for each data point, we resample the ensemble of independent training runs with replacement and recompute the mean, repeating this process to obtain the sampling distribution from which the standard error is extracted.
The uncertainties in the fitting parameters are estimated by the contour $\chi_\nu^2(\sigma, p_{\text{eval},c}) = 1.3\chi_{\nu,\min}^2$ in parameter space as an operational definition, where $\chi_{\nu,\min}^2$ is the global minimum of $\chi_\nu^2$, and the extent of this contour along each axis is taken as the error bar.

\begin{figure}[htbp]
    \centering
    \includegraphics[width=3.4in]{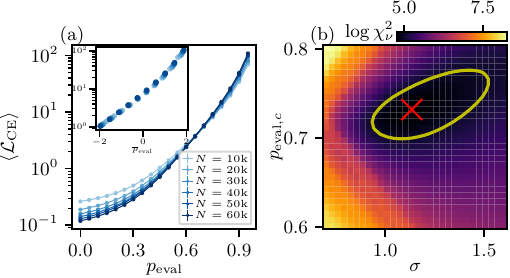}
    \caption{BKT-like finite-size scaling analysis at $p_{\text{train}} = 0.2$. (a) Averaged cross-entropy loss versus $p_{\text{eval}}$ for different training dataset size $N$. Inset: data collapse with $p_{\text{eval},c} = 0.73(4)$ and $\sigma = 1.1(4)$, where $\overline{p}_{\text{eval}} \equiv (p_{\text{eval}} - p_{\text{eval},c}) [\log(N/N_0)]^{1/\sigma}$ is the scaling variable in Eq.~\eqref{eq:bkt}. (b) $\chi_\nu^2(\sigma, p_{\text{eval},c})$ landscape with $\chi_{\nu,\min}^2 = 115$ (cross) and $1.3\,\chi_{\nu,\min}^2$ contour.}
    \label{fig:CE_linecut_BKT_ptrain0.2}
\end{figure}

\begin{figure}[htbp]
    \centering
    \includegraphics[width=3.4in]{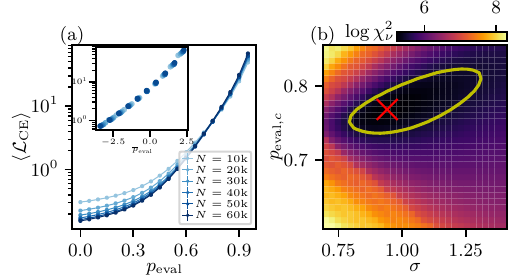}
    \caption{BKT-like finite-size scaling analysis at $p_{\text{train}} = 0.3$. (a) Averaged cross-entropy loss versus $p_{\text{eval}}$ for different training dataset size $N$. Inset: data collapse with $p_{\text{eval},c} = 0.77(4)$ and $\sigma = 0.9(3)$. (b) $\chi_\nu^2(\sigma, p_{\text{eval},c})$ landscape with $\chi_{\nu,\min}^2 = 184$ (cross) and $1.3\,\chi_{\nu,\min}^2$ contour.}
    \label{fig:CE_linecut_BKT_ptrain0.3}
\end{figure}

\begin{figure}[htbp]
    \centering
    \includegraphics[width=3.4in]{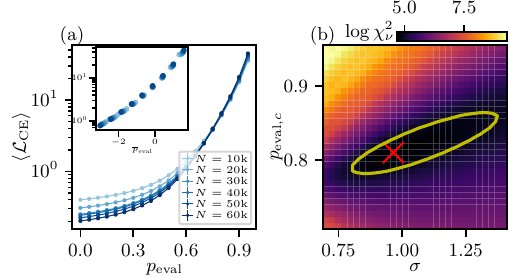}
    \caption{BKT-like finite-size scaling analysis at $p_{\text{train}} = 0.4$. (a) Averaged cross-entropy loss versus $p_{\text{eval}}$ for different training dataset size $N$. Inset: data collapse with $p_{\text{eval},c} = 0.81(4)$ and $\sigma = 1.0(3)$. (b) $\chi_\nu^2(\sigma, p_{\text{eval},c})$ landscape with $\chi_{\nu,\min}^2 = 106$ (cross) and $1.3\,\chi_{\nu,\min}^2$ contour.}
    \label{fig:CE_linecut_BKT_ptrain0.4}
\end{figure}

\begin{figure}[htbp]
    \centering
    \includegraphics[width=3.4in]{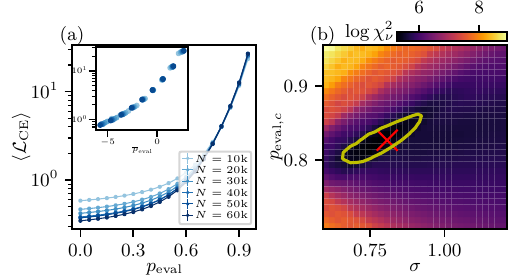}
    \caption{BKT-like finite-size scaling analysis at $p_{\text{train}} = 0.5$. (a) Averaged cross-entropy loss versus $p_{\text{eval}}$ for different training dataset size $N$. Inset: data collapse with $p_{\text{eval},c} = 0.83(3)$ and $\sigma = 0.8(1)$. (b) $\chi_\nu^2(\sigma, p_{\text{eval},c})$ landscape with $\chi_{\nu,\min}^2 = 188$ (cross) and $1.3\,\chi_{\nu,\min}^2$ contour.}
    \label{fig:CE_linecut_BKT_ptrain0.5}
\end{figure}

\para{Example of other data collapse at different $p_{\text{train}}$, showing that it is good for small $p_{\text{train}}$ but becomes worse for large $p_{\text{train}}$}
In Figs.~\ref{fig:CE_linecut_BKT_ptrain0.2}--\ref{fig:CE_linecut_BKT_ptrain0.5}, we present the BKT-like finite-size scaling analysis at $p_{\text{train}} = 0.2$, $0.3$, $0.4$, and $0.5$, complementing the $p_{\text{train}} = 0.1$ case shown in Fig.~\ref{fig:CE_linecut_BKT}. In each case, the cross-entropy loss exhibits a crossing for different training dataset sizes $N$ as a function of $p_{\text{eval}}$, and the data collapses following the ansatz in Eq.~\eqref{eq:bkt}, confirming that the eumentia--dementia transition is generically BKT-like across the phase boundary. The quality of the collapse, however, degrades with increasing $p_{\text{train}}$: $\chi_{\nu,\min}^2$ grows from $26.6$ at $p_{\text{train}} = 0.1$ to values exceeding $100$ for $p_{\text{train}} \geq 0.2$. This deterioration occurs because the critical point $p_{\text{eval},c}$ shifts toward the boundary $p_{\text{eval}} = 1$ at larger $p_{\text{train}}$, compressing the range of available data on the disordered side and making the collapse increasingly constrained.

\section{finite-size scaling using conventional algebraic divergence}\label{app:powerlaw}

\para{Derive the finite-size scaling ansatz for the conventional second order transition, since the previous section already set up the stage so you dont' need to excessive repetition again, just striaght to the point.}
For a conventional second-order transition, the correlation length diverges algebraically,
\begin{equation}
\xi \sim \abs{p - p_c}^{-\nu},
\end{equation}
where $\nu$ is the correlation length exponent. Following the same finite-size scaling argument as in Appendix~\ref{app:bkt}, the condition $\xi \sim L$ gives $\abs{p - p_c} \sim L^{-1/\nu}$, and the scaling variable becomes $(p - p_c)\, L^{1/\nu}$. With $N$ playing the role of $L$, the finite-size scaling ansatz for the cross-entropy loss is
\begin{equation}\label{eq:powerlaw}
\expval{\mathcal{L}_{\text{CE}}} \sim g\left((p_{\text{eval}} - p_{\text{eval},c})\, N^{1/\nu}\right),
\end{equation}
where $g$ is a universal scaling function and $\nu$ and $p_{\text{eval},c}$ are the two fitting parameters. The definition of $\chi_\nu^2$ and the error estimation procedure are identical to those in Appendix~\ref{app:bkt}, with the scaling variable $x_i = (p_{\text{eval},i} - p_{\text{eval},c})\, N_i^{1/\nu}$.

\begin{figure}[htbp]
    \centering
    \includegraphics[width=3.4in]{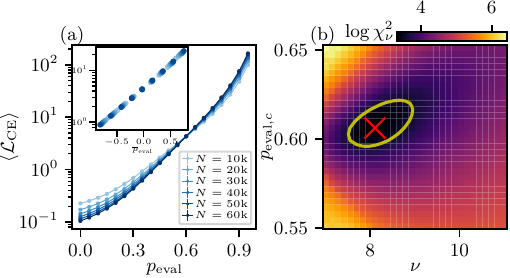}
    \caption{Conventional algebraic finite-size scaling using Eq.~\eqref{eq:powerlaw} at $p_{\text{train}} = 0.1$. The fitting parameters are $\nu$ and $p_{\text{eval},c}$. (a) Averaged cross-entropy loss versus $p_{\text{eval}}$ for different training dataset size $N$. Inset: data collapse with $p_{\text{eval},c} = 0.61(2)$ and $\nu = 8.1(7)$, where $\overline{p}_{\text{eval}} \equiv (p_{\text{eval}} - p_{\text{eval},c})\, N^{1/\nu}$ is the scaling variable in Eq.~\eqref{eq:powerlaw}. (b) $\chi_\nu^2(\nu, p_{\text{eval},c})$ landscape with $\chi_{\nu,\min}^2 = 27.4$ (cross) and $1.3\,\chi_{\nu,\min}^2$ contour.}
    \label{fig:CE_linecut_powerlaw_ptrain0.1}
\end{figure}

\begin{figure}[htbp]
    \centering
    \includegraphics[width=3.4in]{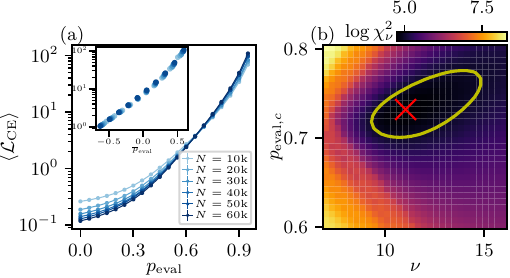}
    \caption{Conventional algebraic finite-size scaling using Eq.~\eqref{eq:powerlaw} at $p_{\text{train}} = 0.2$. (a) Averaged cross-entropy loss versus $p_{\text{eval}}$ for different training dataset size $N$. Inset: data collapse with $p_{\text{eval},c} = 0.73(4)$ and $\nu = 11(3)$. (b) $\chi_\nu^2(\nu, p_{\text{eval},c})$ landscape with $\chi_{\nu,\min}^2 = 114$ (cross) and $1.3\,\chi_{\nu,\min}^2$ contour.}
    \label{fig:CE_linecut_powerlaw_ptrain0.2}
\end{figure}

\begin{figure}[htbp]
    \centering
    \includegraphics[width=3.4in]{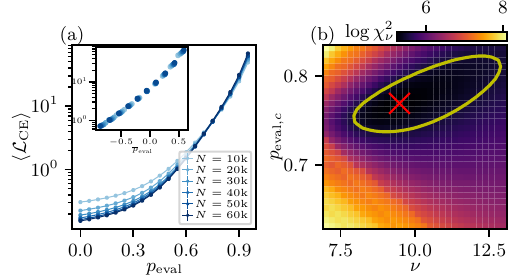}
    \caption{Conventional algebraic finite-size scaling using Eq.~\eqref{eq:powerlaw} at $p_{\text{train}} = 0.3$. (a) Averaged cross-entropy loss versus $p_{\text{eval}}$ for different training dataset size $N$. Inset: data collapse with $p_{\text{eval},c} = 0.77(4)$ and $\nu = 9(2)$. (b) $\chi_\nu^2(\nu, p_{\text{eval},c})$ landscape with $\chi_{\nu,\min}^2 = 185$ (cross) and $1.3\,\chi_{\nu,\min}^2$ contour.}
    \label{fig:CE_linecut_powerlaw_ptrain0.3}
\end{figure}

\begin{figure}[htbp]
    \centering
    \includegraphics[width=3.4in]{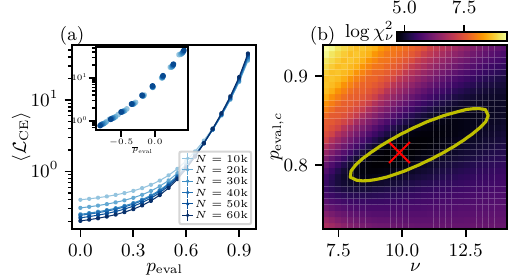}
    \caption{Conventional algebraic finite-size scaling using Eq.~\eqref{eq:powerlaw} at $p_{\text{train}} = 0.4$. (a) Averaged cross-entropy loss versus $p_{\text{eval}}$ for different training dataset size $N$. Inset: data collapse with $p_{\text{eval},c} = 0.81(4)$ and $\nu = 10(3)$. (b) $\chi_\nu^2(\nu, p_{\text{eval},c})$ landscape with $\chi_{\nu,\min}^2 = 106$ (cross) and $1.3\,\chi_{\nu,\min}^2$ contour.}
    \label{fig:CE_linecut_powerlaw_ptrain0.4}
\end{figure}

\begin{figure}[htbp]
    \centering
    \includegraphics[width=3.4in]{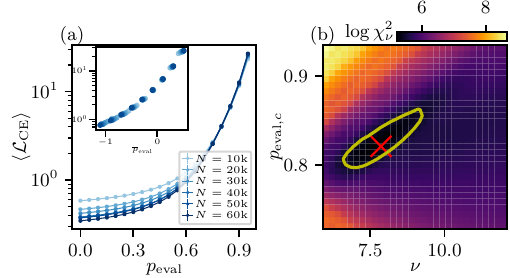}
    \caption{Conventional algebraic finite-size scaling using Eq.~\eqref{eq:powerlaw} at $p_{\text{train}} = 0.5$. (a) Averaged cross-entropy loss versus $p_{\text{eval}}$ for different training dataset size $N$. Inset: data collapse with $p_{\text{eval},c} = 0.82(3)$ and $\nu = 8(1)$. (b) $\chi_\nu^2(\nu, p_{\text{eval},c})$ landscape with $\chi_{\nu,\min}^2 = 181$ (cross) and $1.3\,\chi_{\nu,\min}^2$ contour.}
    \label{fig:CE_linecut_powerlaw_ptrain0.5}
\end{figure}

\para{Present the data collapse for using algebraic divergence again for the same $p_{\text{train}}=0.1$  to 0.5, mention that (1) the power law can also fit with a slightly larger $\chi_\nu^2$  meaning it is slightly worse than the BKT-like transition , echoing line 269;  but the large nu is an implication of BKT, so we are still believe it implies BKT is more likely}
In Figs.~\ref{fig:CE_linecut_powerlaw_ptrain0.1}, \ref{fig:CE_linecut_powerlaw_ptrain0.2}, \ref{fig:CE_linecut_powerlaw_ptrain0.3}, \ref{fig:CE_linecut_powerlaw_ptrain0.4}, and \ref{fig:CE_linecut_powerlaw_ptrain0.5}, we present the conventional algebraic finite-size scaling analysis using Eq.~\eqref{eq:powerlaw} for $p_{\text{train}} = 0.1$ through $0.5$. The data also collapses under the algebraic ansatz, with $\chi_{\nu,\min}^2$ values comparable to those obtained from the BKT-like ansatz (cf.\ Figs.~\ref{fig:CE_linecut_BKT}, \ref{fig:CE_linecut_BKT_ptrain0.2}, \ref{fig:CE_linecut_BKT_ptrain0.3}, \ref{fig:CE_linecut_BKT_ptrain0.4}, and \ref{fig:CE_linecut_BKT_ptrain0.5}). However, the fitted correlation length exponent is consistently large, $\nu \gtrapprox 8$, across all values of $p_{\text{train}}$. Since the BKT stretched exponential is the $\nu \to \infty$ limit of the algebraic divergence, these large values of $\nu$ are themselves consistent with a BKT-like transition and further support the interpretation in Sec.~\ref{sec:bkt}.

\section{MNIST Dataset}\label{app:MNIST}
The MNIST dataset is a widely used benchmark dataset composed of $70000$ grayscale images of handwritten digits ($0-9$), divided into $60000$ training and $10000$ testing samples. The dataset is class-balanced, with an approximately equal number of samples for each of the $10$ digit classes. This near-uniform distribution mitigates class-imbalance effects during training and ensures that classification performance is not biased toward any particular digit. 

Each MNIST image has a spatial resolution of $28\times 28$ pixels, where each pixel is an integer grayscale value from 0 (black) to 255 (white). We rescale all pixel values by dividing by 255, so that they lie in the interval $[0,1]$.

Each image is then flattened into a one-dimensional input vector $\bm{x} \in [0,1]^{784}$, which is fed to the input layer of the FCNN described in Sec.~\ref{sec:fcnn}.

\section{Neuron Dropout}\label{app:dropout}
Neuron dropout \cite{srivastava2014dropout} is a stochastic regularization method in which a random subset of neurons in a layer is temporarily deactivated by setting their post-activation output to $0$. To build intuition, consider a simplified scenario in digit classification. In the absence of dropout, suppose that one neuron becomes highly specialized in detecting vertical lines in the input (e.g., those characteristic of the digit `1'). A neighboring neuron may become heavily dependent on that first neuron to recognize the digit `$1$'. If the first neuron makes a mistake, the entire prediction may fail, and the system becomes fragile. This phenomenon is referred to as `co-adaptation', where neurons develop strong mutual dependencies. 

With dropout, the vertical-line detecting neuron is sometimes turned off. But, the network must still correctly classify the digit. As a result, other neurons are compelled to learn to extract and represent vertical-line features and no individual neuron becomes indispensable. Thus, dropout suppresses co-adaptations, reduces overfitting and promotes better model stability, which can help to generalize effectively to unseen data during the testing phase. 

In our model architecture (Sec.~\ref{sec:fcnn}), each neuron $i$ in the hidden layers ($l=1,\dots,L$) is independently dropped with probability $p$, setting its post-activation output $h_i$ to zero, or retained with probability $1-p$. This introduces multiplicative noise into the network activations.

In practice, the dropout is implemented as \emph{inverted dropout}, where the surviving activations are rescaled whenever dropout is active to preserve their expected magnitude, given by
\begin{equation}
\tilde{h}_i = \frac{m_i}{1-p} h_i .
\end{equation}
where $m_i \in \{0,1\}$ is a Bernoulli random variable, representing the dropout mask for neuron $i$. With this normalization, one has $\mathbb{E}[\tilde{h}_i] = h_i$, ensuring that the typical scale of activations remains unchanged between training and testing. During training, we apply a dropout rate $p=p_{\text{train}}$ and optimize the model parameters $\vec{\theta}$. In every forward pass of the training, a different dropout mask is applied to the hidden layer neurons. After training, the learned parameters are fixed. During the testing phase, the dropout is applied with the rate $p=p_{\text{eval}}$, and the performance of the resulting model is quantified by the average classification accuracy and cross-entropy loss in the test dataset.

\end{document}